\documentclass[aps,prd,floatfix,nofootinbib,superscriptaddress,preprint]{revtex4-1}
\usepackage{graphicx}
\usepackage{amsmath,amssymb,amsfonts}
\usepackage{mathrsfs}
\usepackage{xcolor}
\usepackage{appendix}

\allowdisplaybreaks

\newcommand\vecl[1]{{\bf{#1}}}        			  		 	 % vector letter
\newcommand\vecs[1]{\boldsymbol{#1}}    				   	 % vector symbol
\newcommand\nn[0]{\nonumber}					 			 % no number
\newcommand\ovbb[0]{0\nu\beta\beta}			 			 % 0vbb
\newcommand\vvbb[0]{2\nu\beta\beta}			 			 % 2vbb
\usepackage{hyperref}

\definecolor{urls}{HTML}{cc003e}
\definecolor{links}{HTML}{00A4CC}

\hypersetup{
    colorlinks,
    linkcolor={links!80!black},
    citecolor={urls},
    urlcolor={urls}
}

\newcommand{\AddrUCL}{Department of Physics and Astronomy, University College London,\\London WC1E 6BT, United Kingdom}

\newcommand{\AddrMPIK}{Max-Planck-Institut f\"ur Kernphysik, Saupfercheckweg 1, 69117 Heidelberg, Germany}

\newcommand{\AddrBLTP}{BLTP, JINR, 141980 Dubna, Russia}

\newcommand{\AddrComenius}{Comenius University, Mlynsk\'a dolina F1, SK–842 48 Bratislava, Slovakia}

\newcommand{\AddrIEAP}{IEAP CTU, 128–00 Prague, Czech Republic}

\def\suppmat{Appendix}

\begin{document}
%!TEX root = ./paper-arxiv.tex

\title{Searching for New Physics in Two-Neutrino Double Beta Decay}

\author{Frank F. Deppisch} 
\email{f.deppisch@ucl.ac.uk}\affiliation{\AddrUCL}
\author{Lukas Graf}
\email{lukas.graf@mpi-hd.mpg.de}\affiliation{\AddrMPIK}
\author{Fedor {\v S}imkovic}
\email{fedor.simkovic@fmph.uniba.sk}\affiliation{\AddrBLTP}\affiliation{\AddrComenius}\affiliation{\AddrIEAP}

\begin{abstract}
\noindent 
Motivated by non-zero neutrino masses and the possibility of New Physics discovery, a number of experiments search for neutrinoless double beta decay. While hunting for this hypothetical nuclear process, a significant amount of two-neutrino double beta decay data have become available. Although these events are regarded and studied mostly as the background of neutrinoless double beta decay, they can also be used to probe physics beyond the Standard Model. In this paper we show how the presence of right-handed leptonic currents would affect the energy distribution and angular correlation of the outgoing electrons in two-neutrino double beta decay. Consequently, we estimate constraints imposed by currently available data on the existence of right-handed neutrino interactions without having to assume their nature. In this way our results complement the bounds coming from the non-observation of neutrinoless double beta decay as they limit also the exotic interactions of Dirac neutrinos. We perform a detailed calculation of two-neutrino double beta decay under the presence of exotic (axial-)vector currents and we demonstrate that current experimental searches can be competitive to existing limits.
\end{abstract}

\maketitle

%--------------------------------------------------------------------------------
\section{Introduction}
\label{sec:intro}
%--------------------------------------------------------------------------------
Double beta decay processes are sensitive probes of physics beyond the Standard Model (SM). The SM process of two-neutrino double beta ($2\nu\beta\beta$) decay is among the rarest processes ever observed with half lives of order $T_{1/2}^{2\nu\beta\beta} \sim 10^{19}~\text{yr}$ and longer \cite{Barabash:2019}. Neutrinoless double beta ($0\nu\beta\beta$) decay, with no observation of any missing energy, is clearly the most important mode beyond the SM as it probes the Majorana nature and mass $m_\nu$ of light neutrinos, with current experiments sensitive as $T_{1/2}^{0\nu\beta\beta} \sim (0.1~\text{eV}/m_\nu)^2 \times 10^{26}~\text{y}$. In general, it is a crucial test for any New Physics scenario that violates lepton number by two units~\cite{Deppisch:2012nb, Graf:2018ozy, Cirigliano:2018yza}.

While $0\nu\beta\beta$ decay is the key process, experimental searches for this decay also provide a detailed measurement of the $2\nu\beta\beta$ decay rate and spectrum in several isotopes. For example, Kamland-Zen measures the $2\nu\beta\beta$ decay spectrum in $^{136}$Xe with a high statistics \cite{KamLAND-Zen:2019imh}, but can only do with respect to the sum of energies of the two electrons emitted. On the other hand, the NEMO-3 experiment with the technology to track individual electrons can measure the individual electron energy spectra and the opening angle between the two electrons. This has yielded detailed measurements of the $2\nu\beta\beta$ decay spectra of $^{96}$Zr \cite{Argyriades:2009ph}, $^{150}$Nd \cite{Arnold:2016qyg}, $^{48}$Ca \cite{Arnold:2016ezh}, $^{82}$Se \cite{Arnold:2018tmo} and especially $^{100}$Mo \cite{NEMO-3:2019gwo}, the latter with a very high statistics containing $\approx 5\times 10^{5}$ $2\nu\beta\beta$ decay events. Such measurements are important for the interpretation of $0\nu\beta\beta$ decay searches as it can shed light on the value of the effective axial coupling $g_A$ \cite{Simkovic:2018rdz}. 

The high precision of $2\nu\beta\beta$ decay measurements, expected to continue as the experimental exposures are increased to push the sensitivity of $0\nu\beta\beta$ decay searches, begs the question whether $2\nu\beta\beta$ decay events can be directly used to search for New Physics beyond the SM. This is the focus of this work. We model such new physics effects through effective charged-current operators of the form $\epsilon G_F (\bar e \mathcal{O}_1 \nu)(\bar u\mathcal{O}_2 d)$ with Lorentz structures $\mathcal{O}_1$, $\mathcal{O}_2$ other than the SM $V-A$ type. Here, the Fermi constant $G_F$ is introduced and the small dimensionless coupling $\epsilon$ encapsulates the New Physics effects.

Exotic charged-current operators of the above form are being searched for in nuclear, neutron $\beta$ and pion decays as well as collider searches \cite{Gonzalez-Alonso:2018omy}, giving rise to limits of the order $\epsilon \lesssim 10^{-4} - 10^{-1}$, depending on the Lorentz structure and chirality of the fields involved. In this paper, we will specifically concentrate on exotic operators containing right-handed (RH) vector lepton currents. Such operators prove difficult to constrain as interference with the SM contribution is suppressed by the light neutrino masses. They are nevertheless of strong theoretical interest as their observation, along with the non-observation of lepton number violation would indicate that neutrinos are not Majorana fermions. This is because RH currents with neutrinos but in the absence of a sterile neutrino state would necessarily violate lepton number. In this work, we will show that the existing $2\nu\beta\beta$ data from the NEMO-3 experiment may set the most stringent limits on such operators which are currently only weakly constrained at the 6\% level \cite{Gonzalez-Alonso:2018omy}. We thus describe a novel probe of the fundamental nature of weak interactions and the properties of neutrinos.

%--------------------------------------------------------------------------------
\section{Exotic Charged-Current Interactions}
\label{sec:eft}
%--------------------------------------------------------------------------------
%
\begin{figure*}[t!]
	\centering
	\includegraphics[scale=0.9]{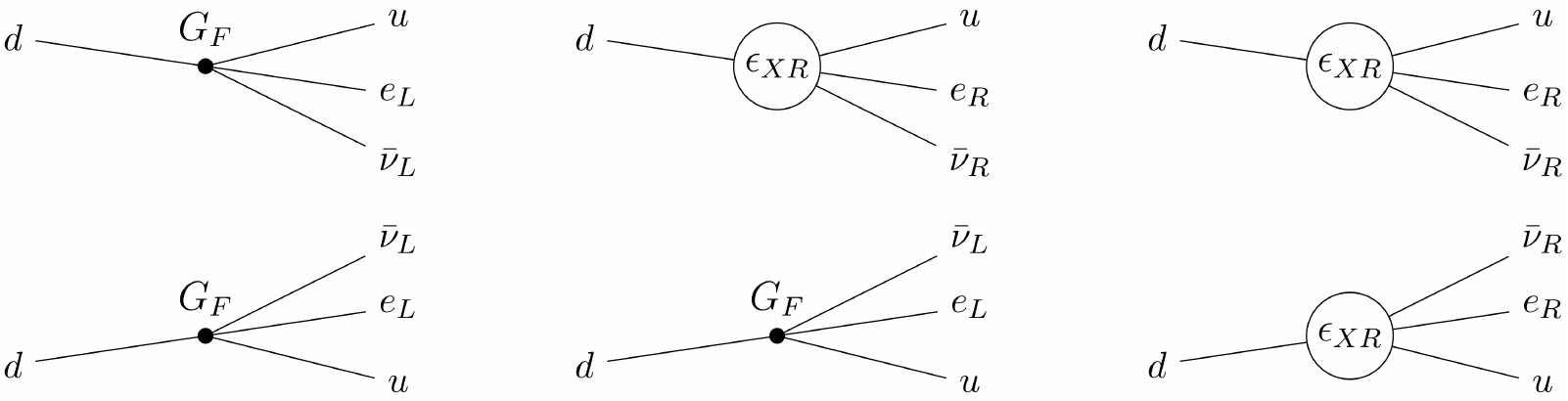}
	\caption{Feynman diagrams for ordinary $2\nu\beta\beta$ decay via the second-order transition through the SM $V-A$ interaction with strength given by the Fermi constant $G_F$~(left),  a transition involving one exotic interaction $\epsilon_{XR}G_F$ with a $V+A$ lepton current of the form $(\bar e_R \mathcal{O}_1 \nu)(\bar u\mathcal{O}_2 d)$~(center) and a second-order transition through the same exotic interaction~(right).
}
\label{fig:diagram} 
\end{figure*}

We are interested in processes where right- and left-handed electrons are emitted considering only $(V+A)$ and $(V-A)$ currents. The effective Lagrangian is written as
\begin{align}
\label{eq:lagrangian}
	\mathcal{L} = 
	\frac{G_F\cos\theta_C}{\sqrt{2}}\left(
	  (1+\delta_\text{SM} + \epsilon_{LL}) j^\mu_L J^{\phantom{\mu}}_{L\mu} 
	                      + \epsilon_{RL}  j_L^\mu J^{\phantom{\mu}}_{R\mu}
		                  + \epsilon_{LR}  j_R^\mu J^{\phantom{\mu}}_{L\mu}
		                  + \epsilon_{RR}  j_R^\mu J^{\phantom{\mu}}_{R\mu}
	\right) + \text{h.c.},
\end{align}
with the tree-level Fermi constant $G_F$, the Cabbibo angle $\theta_C$, and the leptonic and hadronic currents $j_{L,R}^\mu = \bar e\gamma^\mu(1\mp\gamma_5)\nu$ and $J^\mu_{L,R} = \bar u\gamma^\mu(1\mp\gamma_5)d$, respectively. The SM electroweak radiative corrections are encoded in $\delta_{SM}$ and the $\epsilon_{XY}$ encapsulate new physics effects. We here concentrate on the latter two operators with RH lepton currents as they are expected to change the $2\nu\beta\beta$ decay kinematic spectra more significantly. Extensions of the above set of operators can be considered; for example, currents other than vector and axial-vector can be included \cite{Cirigliano:2013} and further, exotic particles may participate~\cite{Cepedello:2018zvr}. 

In Eq.~\eqref{eq:lagrangian}, $\nu$ is a 4-spinor field of the light electron neutrino, either defined by $\nu = \nu_L + \nu_L^c$ (i.e. a Majorana spinor constructed from the SM active left-handed neutrino $\nu_L$ and its charge-conjugate) or $\nu = \nu_L + \nu_R$ (a Dirac spinor constructed from the SM $\nu_L$ and a new SM-sterile RH neutrino $\nu_R$). Whether the light neutrinos are of Majorana or Dirac type and whether total lepton number is broken or conserved is of crucial importance for an underlying model but as far as the effective interactions in Eq.~\eqref{eq:lagrangian} are concerned, this does not play a role in our calculations. If the neutrino in Eq.~\eqref{eq:lagrangian} is a Majorana particle, the operators associated with $\epsilon_{LR}$ and $\epsilon_{RR}$ violate total lepton number by two units and they will give rise to extra contributions to $0\nu\beta\beta$ decay \cite{Doi:1982dn}. In this case, severe limits are set by $0\nu\beta\beta$ decay searches of the order $\epsilon_{LR} \lesssim 3\times 10^{-9}$, $\epsilon_{RR}\lesssim 6\times 10^{-7}$ \cite{Deppisch:2012nb}. On the other hand, if there exists a sterile neutrino Weyl state $\nu_R$ that combines with $\nu_L$ to form a Dirac neutrino, the RH current interactions in Eq.~\eqref{eq:lagrangian} do not necessarily violate lepton number which, in fact, can remain an unbroken symmetry of the underlying model. For example, such effective interactions can emerge in Left-Right symmetric models (LRSMs) \cite{pati:1974yy} with unbroken lepton number \cite{Bolton:2019bou}. The observation of the effect of RH neutrino operators without the observation of lepton number violation would thus strongly suggest that neutrinos are Dirac fermions.

The most stringent direct limits on the above operators for process energies $E\approx$~MeV are set by fitting experimental results of neutron and various nuclear single $\beta$ decays, $\epsilon_{LL}, \epsilon_{RL} \leq 5\times 10^{-4}$, $\epsilon_{LR}, \epsilon_{RR}\leq 6\times 10^{-2}$ \cite{Cirigliano:2013, Gonzalez-Alonso:2018omy}. The limits on the RH lepton currents are much less severe due to the absence of an interference with the SM contribution. Searches at the Large Hadron Collider (LHC) for single electron and missing energy signatures \cite{Khachatryan:2014tva}, $pp\to eX + \text{MET}$, may also be used to constrain the above operators, $\epsilon_{LL} \lesssim 4.5\times 10^{-3}$, $\epsilon_{RR} \lesssim 2.2\times 10^{-3}$ \cite{Gonzalez-Alonso:2013uqa}. While the constraints are stringent and the sensitivity is expected to improve to $\epsilon_{LL} \approx 10^{-5}$~\cite{Greljo:2017vvb}, the LHC operates at a much higher energy and the effective operator analysis is only applicable if the new physics mediators integrated out are much heavier than this. More model-dependent limits can also be set by direct searches for RH current mediators at the LHC \cite{Aaboud:2018spl}, from considerations of sterile neutrino thermalization and the resulting increase of the effective number of light degrees of freedom in the early universe and supernova cooling. The associated new physics scales probed range between $\Lambda \approx 5 - 20$~TeV, corresponding to $\epsilon_{XY} \approx 5\times 10^{-4} - 5\times 10^{-5}$. An indirect limit on $\epsilon_{LR}$ can be set from the fact that the associated operator contributes to the Dirac neutrino mass at the second loop order \cite{Prezeau:2004md}. Using current direct neutrino mass bounds this results in $\epsilon_{LR} \lesssim 10^{-2}$ \cite{Vos:2015eba}. Especially the direct limit $\epsilon_{LR} \leq 6\times 10^{-2}$ is rather feeble and motivates the need to probe for admixtures of exotic currents in the SM Fermi interaction. While underlying scenarios are expected to trigger the other, better constrained operators as well, it is not difficult to envision cases where $\epsilon_{LR}$ or $\epsilon_{RR}$ are dominant. For example, in LRSMs, the operator associated with $\epsilon_{RL}$ is mediated at lowest order by the SM $W$ boson and involves the mixing $\theta_{LR}$ with an exotic $W_R$ boson. This mixing is a priori unrelated to the $m_{W_R}$ scale and $\epsilon_{RL}$ can thus be suppressed compared to $\epsilon_{RR}$ if $\theta_{LR}$ is small. It is also not difficult to think of extensions of the minimal LRSM where exotic copies of quarks are charged under the LRSM $SU(2)_R$ but not the SM quarks. The exotic quarks instead mix with the SM quarks and the latter will inherit a suppressed RH current, suppressing $\epsilon_{RL}$ with respect to $\epsilon_{LR}$.

%--------------------------------------------------------------------------------
\section{Decay Rate and Distributions}
\label{sec:calculation}
%--------------------------------------------------------------------------------
We have calculated the differential rate of $\vvbb$ decay under the presence of the exotic interactions in Eq.~\eqref{eq:lagrangian}. Because $\vvbb$ decay is possible in the SM, arising in second order perturbation theory of the first term in Eq.~\eqref{eq:lagrangian}, interference between SM and exotic contributions is in principle possible. In general, the amplitude of  $\vvbb$ decay is calculated as a coherent sum of the Feynman diagrams in Fig.~\ref{fig:diagram}. To lowest order in $\epsilon_{XR}$, exotic effects occur from the interference of the SM diagram Fig.~\ref{fig:diagram}~(left) and Fig.~\ref{fig:diagram}~(center). Due to the RH nature of the exotic current, such an interference is helicity suppressed by the masses of the emitted electron and neutrino as $m_e m_\nu / Q^2$, with the $\vvbb$ decay energy release $Q$. For light eV-scale neutrinos it is thus utterly negligible.\footnote{This is not necessarily the case if currents other than $V\pm A$ vector currents are considered in Eq.~\eqref{eq:lagrangian}.} Contributions to second-order $\propto \epsilon_{XR}^2$ come from the center diagram and the interference of the SM contribution (left) with the second-order exotic diagram (right). The latter is suppressed even more strongly by the neutrino mass and thus negligible. To lowest order in the exotic coupling, the squared matrix element for ground state to ground state $\vvbb$ transition can thus be written as the incoherent sum
\begin{align}
	|\mathcal{R}^{2\nu}|^2 = |\mathcal{R}^{2\nu}_{\rm SM}|^2 + |\epsilon_{XR}|^2 |\mathcal{R}^{2\nu}_{\epsilon}|^2,
\end{align}
where $\mathcal{R}^{2\nu}_{\rm SM}$ is the matrix element for SM $\vvbb$ decay and $\mathcal{R}^{2\nu}_{\epsilon}$ is the exotic contribution. As discussed in detail in the \suppmat~\cite{SuppMat}, the latter may be expressed as
\nocite{Doi:1985dx, Haxton:1985am, Gysbers:2019uyb}
\begin{align}
  \label{eq:totM1}
  \mathcal{R}^{2\nu}_{\epsilon}  &= i \left(\frac{1}{\sqrt{2}}\right)^2 \left(\frac{G_F\cos\theta_W}{\sqrt{2}}\right)^2
        [1-\text{P}(e_1,e_2)] [1-\text{P}(\bar\nu_1,\bar\nu_2)] \nn \\
	&\times \left[~\overline{\psi}(p_{e_1})\gamma^\mu(1+\gamma_5)\psi^{c}(p_{\bar\nu_1})~
          \overline{\psi}(p_{e_2})\gamma^\nu(1-\gamma_5)\psi^{c}(p_{\bar\nu_2}) \right. \nn \\
	& ~~+ \left.\overline{\psi}(p_{e_1})\gamma^\nu(1-\gamma_5)\psi^{c}(p_{\bar\nu_1})~
          \overline{\psi}(p_{e_2})\gamma^\mu(1+\gamma_5)\psi^{c}(p_{\bar\nu_2}) \right] \nn \\
	&\times \left( g_{\mu 0}g_{\nu 0}~ g_V^2 M_F
        \mp\frac{1}{3} g_{\mu k} g_{\nu k}~ g_A^2 M_{GT}\right),
\end{align}
where $\psi(p_f)$ is the wave function of the emitted fermion $f$ with momentum $p_f$ and we consider here the commonly used approximation of the $S_{1/2}$ wave evaluated at the nuclear surface. The nuclear matrix elements between the initial $0_i^+$, the intermediate $0^+_n$ ($1^+_n$) and the final $0^+_f$ states of the nucleus are generally of Fermi (Gamow-Teller) type with the associated nucleon-level vector (effective axial-vector) coupling $g_V$ ($g_A$),
\begin{align}
	M_F &= \sum_n
		\frac{\langle 0^+_f|\sum_j\tau^+_j |0^+_n\rangle
		\langle 0^+_n|\sum_k\tau^+_k |0^+_i\rangle}
                {\Delta E_n(0_n^+) + E_{e_2}+E_{\bar\nu_2}},\nonumber\\
	M_{GT} &= \sum_n
		\frac{\langle 0^+_f|\sum_j\tau^+_j {\boldsymbol{\sigma}}_j|1^+_n\rangle \cdot
		\langle 1^+_n|\sum_k\tau^+_k {\boldsymbol{\sigma}}_k | 0^+_i\rangle}
                {\Delta E_n(1_n^+) + E_{e_2}+E_{\bar\nu_2}}. 
\end{align}
The summations are over all intermediate $0^+_n,~1^+_n$ states and all nucleons $j,k$ inside the nucleus where $\tau^+_{j,k}$ is the isospin-raising operator transforming a neutron into a proton and $\vecs{\sigma}_{j,k}$ represents the nucleon spin operator. Assuming isospin invariance, the Fermi matrix elements vanish. The energy denominators arise due to the second-order nature of the above matrix element where $\Delta E_n(J^\pi_n) = E_n(J^\pi_n) - E_i$ ($J^\pi_n = 0^+_n$ and $1^+_n$) are the energies of the intermediate nuclear states with respect to the initial ground state. Overall energy conservation is implied, $E_i = E_f + E_{e_1} + E_{e_2} + E_{\bar\nu_1} + E_{\bar\nu_2}$, and, as indicated by the particle exchange operator P$(a,b)$, the matrix element is anti-symmetrized with respect to the exchange of the identical electrons and antineutrinos (the corresponding anti-symmetrization over the nucleons is implicitly included in the nuclear states).

Following Ref.~\cite{Simkovic:2018rdz}, the calculation of the $\vvbb$ decay rate and distributions is detailed in the \suppmat. We use nuclear matrix elements in the QRPA formalism from Ref.~\cite{Simkovic:2018rdz} assuming isospin invariance with $M_F = 0$ and including higher order corrections from the effect of the final state lepton energies. Because of $M_F = 0$ and negligible SM -- exotic interference effects, the calculations for $\epsilon_{LR}$ and $\epsilon_{RR}$ are identical; both cases yield the same rates and distributions. As a result, we calculate the full differential $\vvbb$ decay rate in a given $0^+ \to 0^+$ double beta decaying isotope with respect to the two electron energies $m_e \leq E_{e_1,e_2} \leq Q + m_e$ and the angle $0 \leq \theta \leq \pi$ between the emitted electrons, which may be written as
\begin{align}
\label{eq:fullydifferential}
	\frac{d\Gamma^{2\nu}}{dE_{e_1}dE_{e_2}d\!\cos\theta} = \frac{\Gamma^{2\nu}}{2} \frac{d\Gamma^{2\nu}_\text{norm}}{dE_{e_1}dE_{e_2}}
	\left(1 + \kappa^{2\nu}(E_{e_1},E_{e_2})\cos\theta\right).
\end{align}
Because interference effects between the SM and the RH current diagram are negligible, the differential rate is simply the incoherent sum of both. In the \suppmat{} we describe in detail the calculation of the above differential decay rate and the derived energy distributions, angular correlations and total rate. Specifically, for $^{100}$Mo the total decay rate $\Gamma^{2\nu} = \ln 2/T_{1/2}^{2\nu}$ associated with the $\vvbb$ half-life $T_{1/2}^{2\nu}$ may be approximated as $\Gamma^{2\nu} \approx \Gamma^{2\nu}_{\rm SM}(1 + 6.11\,\epsilon^2_{XR})$, where $\Gamma^{2\nu}_\text{SM}$ is the SM rate. The experimentally accessible kinematic information is contained in the normalized double-differential energy distribution $d\Gamma^{2\nu}_\text{norm}/(dE_{e_1}dE_{e_2})$ and the energy-dependent angular correlation $-1 < \kappa^{2\nu}(E_{e_1},E_{e_2}) < 1$. The latter determines whether the two electrons are preferably emitted back-to-back ($\kappa^{2\nu} \approx -1$), in the same direction ($\kappa^{2\nu} \approx 1$) or in intermediate configurations.

\begin{figure}[t!]
	\centering
	\includegraphics[width=0.49\textwidth]{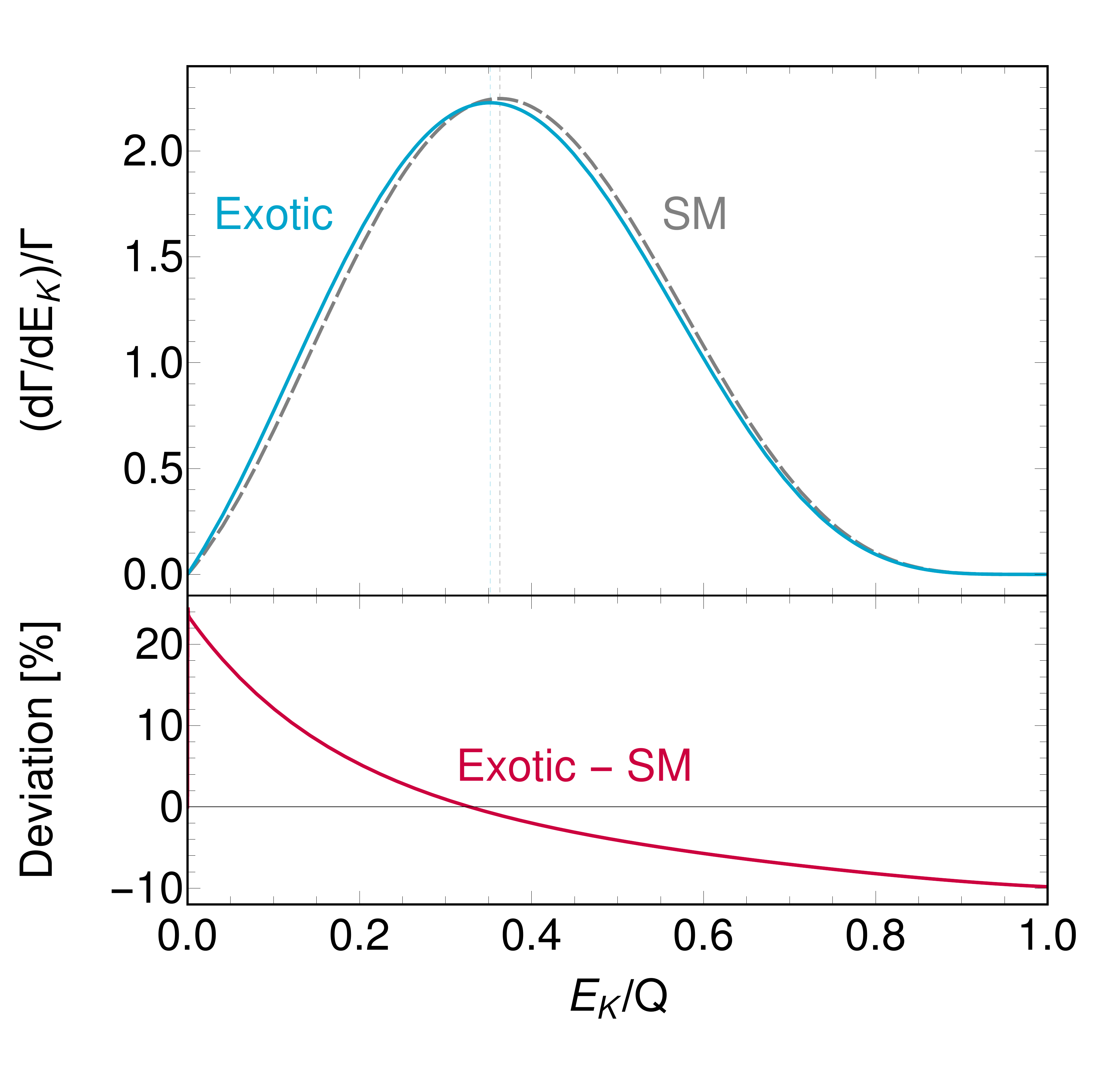}
	\includegraphics[width=0.49\textwidth]{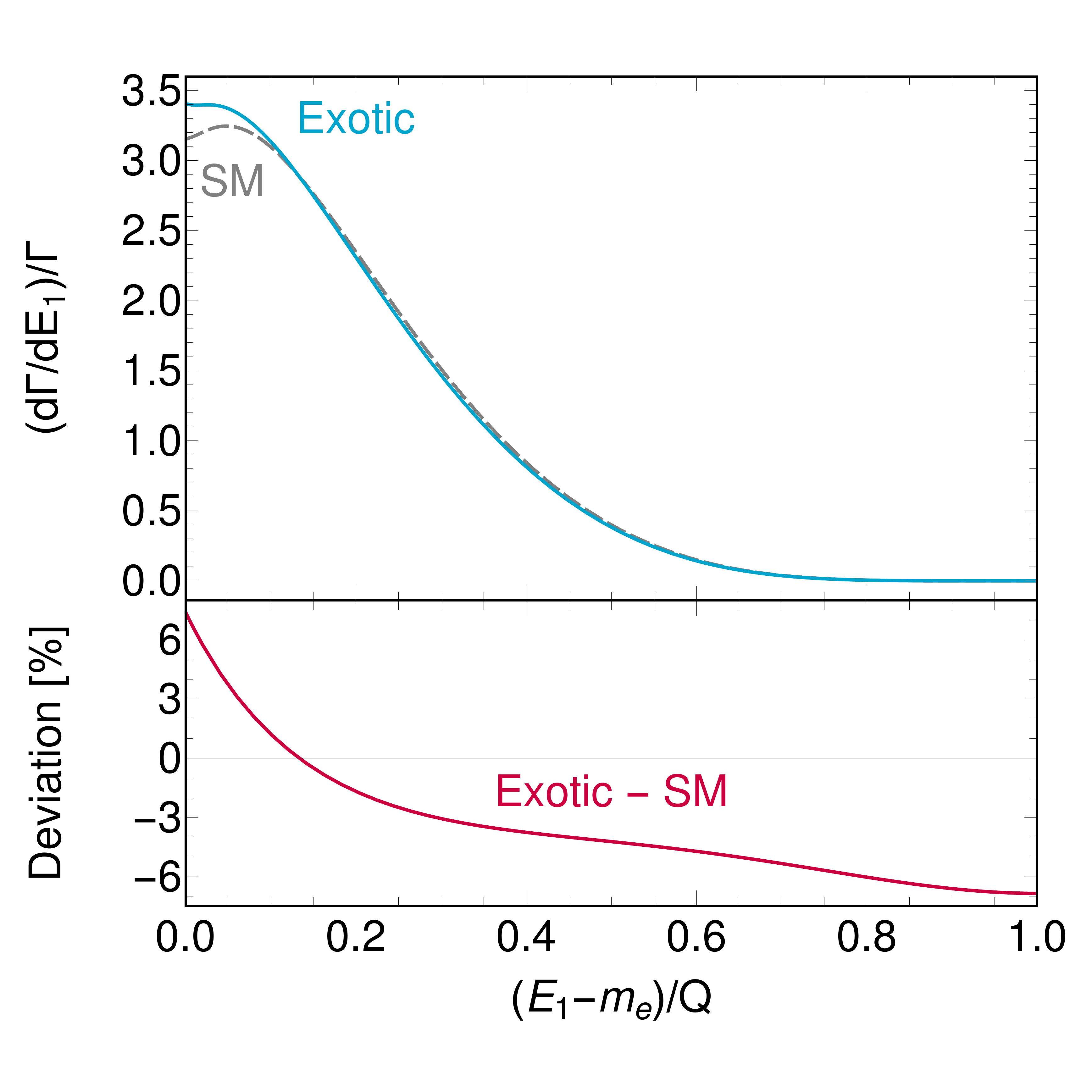}
	\caption{Left: Normalized $\vvbb$ decay distributions with respect to the total kinetic energy $E_K = E_{e_1} + E_{e_2} - 2m_e$ of the emitted electrons for standard $\vvbb$ decay through SM $V-A$ currents (dashed) and a pure RH lepton current (solid). Right: Normalized $\vvbb$ decay distributions with respect to the energy of a single electron in the same scenarios. Both plots are for the isotope $^{100}$Mo and the energies are normalized to the $Q$ value. The bottom panels show the relative deviation of the exotic distribution from the SM case.}
	\label{fig:EnergyDistros1D}
\end{figure}
Given the uncertainties in nuclear matrix elements, the change of the total decay rate due to the presence of a RH current contribution is not expected to be measurable. Instead, differences in spectral shape of either the energy or angular distributions may be more sensitive. All double beta decay experiments measure the spectrum of events with respect to the sum of the electron kinetic energies, $E_K = E_{e_1} + E_{e_2} - 2m_e$. For $^{100}$Mo, it is shown in Fig.~\ref{fig:EnergyDistros1D}~(left), comparing the $\vvbb$ decay distributions in the SM case (dashed) and for the exotic leptonic RH current operators (solid). The deviation is sizeable leading to a shift of the spectrum to smaller energies and a flatter profile near the endpoint~$E_K / Q = 1$. We find that relative deviations of the order of 10\% for small energies and near the endpoint are expected to occur. In experiments that are able to track and measure the individual electrons, such as NEMO-3 and SuperNEMO, the full doubly-differential energy spectrum is in principle measurable. Alternatively, the spectrum with respect to the kinetic energy of a single electron is shown in Fig.~\ref{fig:EnergyDistros1D}~(right). It helps explain the shift of the energy sum spectrum in the exotic case as each electron receives on average less energy than in the SM.

This behaviour can be traced to the kinematic differences. In the presence of a RH lepton current in $\vvbb$ decay, the electrons are preferably emitted collinearly and the electron energy-dependent correlation factor is always $\kappa^{2\nu}_{\epsilon} > 0$ whereas in the SM case the electrons are preferably emitted back-to-back with $\kappa^{2\nu}_{\rm SM} < 0$. This behaviour can be understood from angular momentum considerations when the two electrons are produced with opposite dominant helicities. Integrating Eq.~\eqref{eq:fullydifferential} over the electron energies one arrives at the angular distribution,
\begin{align}
\label{eq:angular-distribution}
	\frac{d\Gamma^{2\nu}}{d\!\cos\theta} = 
	\frac{\Gamma^{2\nu}}{2} \left(1 + K^{2\nu}\cos\theta\right),
\end{align}
with the angular correlation factor $K^{2\nu}$. For $^{100}$Mo, we calculate $K^{2\nu}_{\rm SM} = -0.626$ in the SM and $K^{2\nu}_{\epsilon} = +0.370$ for the exotic contribution. This deviation is clearly the most striking consequence of a RH lepton current on $\vvbb$ decay. For small $\epsilon_{XR} \ll 1$, the angular correlation factor $K^{2\nu}$ can be expanded as 
\begin{align}
	K^{2\nu} = K^{2\nu}_\text{SM} + \alpha\,\epsilon_{XR}^2 + \mathcal{O}(\epsilon_{XR}^4).
\end{align}
For $^{100}$Mo, the coefficient $\alpha$ turns out to be $\alpha = 6.078$. Despite the small correction expected if $\epsilon_{XR} \approx 10^{-2}$ as indicated in current bounds, searches for $\vvbb$ decay can be sensitive in this regime. A simple signature is to look for the forward-backward asymmetry $A^{2\nu}_\theta$, comparing the number of $\vvbb$ decay events with the electrons being emitted with a relative angle $\theta < \pi/2$ and $\theta > \pi/2$,
\begin{align}
\label{eq:forward-backward}
	A^{2\nu}_\theta = \frac{N_{\theta>\pi/2} - N_{\theta<\pi/2}}{N_{\theta>\pi/2} + N_{\theta<\pi/2}} = \frac{1}{2}K^{2\nu}.
\end{align}
As shown, the asymmetry is simply related to the angular correlation factor $K^{2\nu}$ and it is clearly independent of the overall $\vvbb$ decay rate. Considering only the statistical error, with $N_\text{events} = 5\times 10^5$ $2\nu\beta\beta$ decay events at NEMO-3, the angular correlation coefficient should be measurable with an uncertainty $K^{2\nu}_\text{SM} = -0.6260 \pm 0.0027$. No significant deviation from this SM expectation should then constrain $\epsilon_{XR} \lesssim 2.7\times 10^{-2}$ at 90\% confidence level. This would already improve on the single $\beta$ decay constraint of $6\times 10^{-2}$ \cite{Cirigliano:2013}. If an experiment such as SuperNEMO were able to achieve an increase in exposure by three orders of magnitude, the expected future sensitivity, scaling as $1/\sqrt{N_{\text{events}}}$, would be $\epsilon_{XR} \lesssim 4.8\times 10^{-3}$. This only gives a very rough order of magnitude estimate and a dedicated experimental analysis is required to verify the sensitivity. For example, at NEMO-3 and SuperNEMO, detector effects will result in a reduced acceptance for small electron angles thus affecting the systematic uncertainty~\cite{NEMO-3:2019gwo, Arnold:2010tu}. We note though that it is not strictly necessary to measure the forward-backward asymmetry in Eq.~\eqref{eq:forward-backward}. Instead, even if only including events with $\cos\theta < 0$, where the majority of $2\nu\beta\beta$ events occur, will allow to fit the angular distribution in Eq.~\eqref{eq:angular-distribution}, albeit with a lower statistical significance. In this back-to-back region with $\cos\theta \lesssim 0$ the existing NEMO-3 data is well within the statistical fluctuations \cite{NEMO-3:2019gwo}. As a very rough but conservative estimate, dropping half of the events will give a dataset limited by statistics. This would weaken our estimated sensitivity by a factor of $\sqrt{2}$.

We must also consider the theoretical uncertainty in predicting the angular correlation. Our results were calculated within the nuclear structure framework of the pn-QRPA with partial isospin restoration \cite{Simkovic:2018rdz}. We consider the three main sources of theoretical errors: 
\begin{enumerate}
\item[(i)] The spectrum of intermediate nuclear states as calculated in different nuclear structure models has a small but potentially significant impact on the external lepton phase space and thus the angular correlation. To conservatively model this, we vary the effective axial coupling $g_A$ between $g_A = 0.8$ and $g_A = 1.269$ as described in Ref.~\cite{Simkovic:2018rdz}. This drastically changes the associated $^{100}$Mo matrix element by a factor of 2.6 and thus the decay rate by a factor of $\approx 6.8$ but the SM angular correlation changes only as $K^{2\nu}_\text{SM} = -0.6260\pm 0.0030$. Thus the very conservatively estimated theoretical error is of the same order as the current statistical error. It will be crucial to reduce it to match the improved future statistical uncertainty, though. 

\item[(ii)] In Eq.~\eqref{eq:lagrangian} we only include the fundamental parton-level interactions and we neglect higher-order nuclear currents, namely the induced weak magnetism and pseudo-scalar currents. Their dominant effect on the amplitude will occur in the interference between the latter and the axial-vector nuclear current, which is suppressed by $\lesssim Q^2 / (3 m_\pi^2) \approx 2\times 10^{-4}$ \cite{Tomoda:1990rs}, where $m_\pi$ is the pion mass. This results in a currently negligible correction.

\item[(iii)] For simplicity, we analytically treat the outgoing electron wave functions in the so called Fermi approximation. The proper Coulomb interaction with the nucleus and the electron cloud can be calculated numerically \cite{Kotila:2012zza}, leading to a 15\% correction in the resulting phase space factor but only a negligible shift in the SM angular correlation of $0 < \Delta K_\text{SM}^{2\nu} \lesssim 10^{-4}$ \cite{Kotila:2012zza}.
\end{enumerate}

As can be seen in Fig.~\ref{fig:EnergyDistros1D}~(right), the effect of RH currents is similar to that of varying the contribution of intermediate nuclear states as described in \cite{Simkovic:2018rdz}. It exhibits a similar variation for small electron energies near the peak, depending on single state dominance (SSD) vs. higher state dominance (HSD) modelling of the intermediate nuclear state contributions \cite{NEMO-3:2019gwo}. This has the benefit that experimental searches for these effects, such as described in \cite{KamLAND-Zen:2019imh, NEMO-3:2019gwo, Azzolini:2019yib}, could be adapted to our scenario.

%--------------------------------------------------------------------------------
\section{Conclusions}
\label{sec:discussion}
%--------------------------------------------------------------------------------
Nuclear double beta decay with the emission of two neutrinos and nothing else was proposed over 80 years ago \cite{GoeppertMayer:1935qp} as a consequence of the Fermi theory of single $\beta$ decay. Its main role for particle physics has largely been confined to being an irreducible background to the exotic and yet unobserved lepton number violating neutrinoless ($0\nu\beta\beta$) mode. We have demonstrated here, for the first time to our knowledge, that $\vvbb$ decay may be used in its own right as a probe of new physics. Our result shows that searches for deviations in the spectrum of $\vvbb$ decay can be competitive to existing limits. This provides a motivation to utilize the already large set of observed $\vvbb$ decay events to probe exotic scenarios. The number of events will necessarily increase in the future by one to two orders of magnitude, as $0\nu\beta\beta$ decay is being searched for in future experiments.

We have here focussed on the case of effective operators with RH chiral neutrinos where the interference with the SM contributions is negligible due to the suppression by the neutrino mass. The exotic contribution to observables is therefore proportional to the square of the small New Physics parameter. As a result, such operators are comparatively weakly constrained. They still play an important role in our understanding of neutrinos as the RH nature can be accommodated in one of two ways: (i) through the right-chiral part of the SM neutrino as a Majorana fermion in which case the associated operators will also induce the lepton number violating $0\nu\beta\beta$ decay mode at a level that is already ruled out; or (ii) through the presence of a separate RH neutrino state that, while sterile under the SM gauge interactions, participates in exotic interactions beyond the SM. In the latter case, neutrinos are expected to be Dirac fermions and the observation of RH neutrino currents while the lepton number violating $0\nu\beta\beta$ decay is not observed would indicate this scenario.

If other operators such as scalar currents are considered, interference can be sizeable and even larger effects may be seen, although existing limits such as those from single $\beta$ decay are expected to be more restrictive as well. As we have demonstrated in the example of exotic RH vector currents, while the search for $0\nu\beta\beta$ decay and thus the Majorana nature of neutrinos is the main motivation, the properties of the second-order SM process of $2\nu\beta\beta$ decay can also contain potential hints for New Physics.

\begin{acknowledgments}
FFD acknowledges support from the UK Science and Technology Facilities Council (STFC) via a Consolidated Grant (Reference ST/P00072X/1). FFD is grateful to HEPHY Vienna and together with LG to the Comenius University Bratislava where part of the work was completed. FFD would also like to thank Morten Sode for collaboration in the early stages of the project. F\v{S} acknowledges support by the VEGA Grant Agency of the Slovak Republic under Contract No. 1/0607/20 and by the Ministry of Education, Youth and Sports of the Czech Republic under
the INAFYM Grant No. CZ.02.1.01/0.0/0.0/16\_019/0000766.
\end{acknowledgments}

\appendix*
%!TEX root = ./paper-arxiv.tex

%--------------------------------------------------------------------------------
\section*{A. Calculation of Two-Neutrino Double Beta Decay}
\label{app:calc}
%--------------------------------------------------------------------------------
\renewcommand\thefigure{A.\arabic{figure}}  
\setcounter{figure}{0}
\setcounter{equation}{0}

The $\vvbb$ decay rate can be calculated using the expression \cite{Doi:1985dx}
\begin{align}
\label{eq:totdiffrate}
	d\Gamma 
	= 2\pi \delta(E_{e_1}+E_{e_2}+E_{\bar\nu_1}+E_{\bar\nu 2}+E_f-E_i) \sum_\text{spins} |\mathcal{R}^{2\nu}|^2
	~d \Omega_{e_1} ~d\Omega_{e_2} ~d \Omega_{\bar\nu_1} ~d \Omega_{\bar\nu_2},
\end{align}
where $E_i$, $E_f$, $E_{e_i} = \sqrt{p_{e_i}^2 + m^2_e}$ and $E_{\bar\nu_i} = \sqrt{p_{\bar\nu_i}^2 + m^2_{\nu}}$ ($i = 1, 2$) denote the energies of initial and final nuclei, electrons and antineutrinos, respectively. The magnitudes of the associated spatial momenta are $p_{e_i} = |\vecl{p}_{e_i}|$ and $p_{\bar\nu_i} = |\vecl{p}_{\bar\nu_i}|$ and $m_e$ and $m_{\nu}$ denote the electron and neutrino masses. The phase space differentials are $d\Omega_{e_1} = {d^3\mathbf{p}_{e_1}}/(2\pi)^3$, etc.

After integrating over the phase space of the outgoing neutrinos, the resulting differential $\vvbb$ decay rate can be generally written in terms of the energies $0 \leq E_{e_1}$, $E_{e_2} \leq Q + m_e$ of the two outgoing electrons, with $Q = E_i - E_f - 2m_e$, and the angle $0 \leq \theta \leq \pi$ between the electron momenta $\vecl{p}_{e_1}$ and $\vecl{p}_{e_2}$
as \cite{Doi:1985dx}
\begin{align}
\label{eq:diffrate}
	\frac{d\Gamma^{2\nu}}{dE_{e_1} dE_{e_2} d\!\cos\theta} 
	= c_{2\nu} \left(A^{2\nu} 
	+ B^{2\nu} \cos\theta \right) 
	p_{e_1} E_{e_1} p_{e_2} E_{e_2},
\end{align}
where
\begin{align}
\label{eq:constant}
	c_{2\nu} = \frac{G_{\beta}^4m_e^9}{8\pi^7},
\end{align}
with $G_\beta = G_F\cos{\theta_C}$ ($G_F$ is Fermi constant and $\theta_C$ is the Cabbibo angle).

The quantities $A^{2\nu}$ and $B^{2\nu}$ in Eq.~\eqref{eq:diffrate}, generally functions of the electron energies, are determined by integrating over the neutrino phase space
\begin{align}
\label{eq:ABnuint}
	A^{2\nu} &= 
	\int_{m_{\nu}}^{E_i-E_f-E_{e_1}-E_{e_2}} \mathcal{A}^{2\nu}~ p_{\bar\nu_1} 
	E_{\bar\nu_1}~ p_{\bar\nu_2} E_{\bar\nu_2}~ dE_{\bar\nu_1}, \nn\\ 
	B^{2\nu} &= 
	\int_{m_{\nu}}^{E_i-E_f-E_{e_1}-E_{e_2}} \mathcal{B}^{2\nu}~
        p_{\bar\nu_1} E_{\bar\nu_1}~ p_{\bar\nu_2} E_{\bar\nu_2}~ 
        dE_{\bar\nu_1},
\end{align}
where we used $E_{\bar\nu_2} = E_i - E_f - E_{e_1} - E_{e_2} - E_{\bar\nu_1}$ due to energy conservation. In turn, the quantities ${\cal A}^{2\nu}$ and ${\cal B}^{2\nu}$, generally functions of the electron and neutrino energies, are calculated below using the nuclear and leptonic matrix elements. In the context of our calculation, they may be expressed as
\begin{align}
	{\cal A}^{2\nu} 
	&= {\cal A}^{2\nu}_{\rm SM} 
	 + 2\,\text{Re}(\epsilon_{XR}){\cal A}^{2\nu}_{\epsilon\text{SM}} 
	 + |\epsilon_{XR}|^2 {\cal A}^{2\nu}_{\epsilon}, \nn\\
	{\cal B}^{2\nu} 
	&= {\cal B}^{2\nu}_{\rm SM} 
	 + 2\,\text{Re}(\epsilon_{XR}){\cal B}^{2\nu}_{\epsilon\text{SM}} 
	 + |\epsilon_{XR}|^2 {\cal B}^{2\nu}_{\epsilon},
\label{eq:AB}
\end{align}
expanded in terms of the small exotic coupling coefficients $\epsilon_{RX} = \epsilon_{RL}, \epsilon_{RR}$ of the exotic right-handed currents in Eq.~\eqref{eq:lagrangian} of the main text. Here, we assume that only one exotic contribution is present at a given time. In the following, we will also take the exotic coupling coefficient to be real. The zero order terms ${\cal A}^{2\nu}_{\rm SM}$ and ${\cal B}^{2\nu}_{\rm SM}$ correspond to the standard $\vvbb$ decay mechanism, cf. Fig.~\ref{fig:diagram}~(left) in the main text. The terms ${\cal A}^{2\nu}_{\epsilon}$ and ${\cal B}^{2\nu}_{\epsilon}$ quadratic in $\epsilon_{XR}$ arise from the exotic $\vvbb$ decay mechanism involving one right-handed vector lepton current, cf. Fig.~\ref{fig:diagram}~(center)\footnote{There is also a contribution from the interference of the SM diagram and the second-order exotic diagram in Fig.~\ref{fig:diagram} in the main text, but it is negligible due to neutrino mass suppression.}. Finally, the terms ${\cal A}^{2\nu}_{\epsilon\text{SM}}$ and ${\cal B}^{2\nu}_{\epsilon\text{SM}}$ linear in $\epsilon_{XR}$ correspond to the interference between the two mechanisms. Because of the different electron and neutrino chiralities involved in the standard $V-A$ and the exotic $V+A$ currents, the interference is suppressed as $\approx m_\nu/Q$ and for $|\epsilon_{XR}| \gg m_\nu/Q$, the linear terms are negligible. This is certainly the case for the emission of light active neutrinos with $m_\nu \lesssim 0.1$~eV.

In principle, the chirality of the quark current involved in the considered effective interaction ($\epsilon_{LR}$ or $\epsilon_{RR}$) does affect the resulting $\ovbb$ decay contribution. However, this difference would manifest only as an opposite sign of the Gamow-Teller part of the amplitude. Hence, in the well-motivated approximation of a vanishing double Fermi NME, which we will apply later on, the resulting expressions for the decay rate and the angular correlation of the emitted electrons will not depend on chirality of the considered quark current. Thus, our conclusions will be generally applicable to both effective couplings $\epsilon_{LR}$ and $\epsilon_{RR}$, collectively denoted as $\epsilon_{RX}$.

\subsection*{A.1 First-order contribution in the exotic coupling}
We here describe the calculation of $\vvbb$ under the presence of exotic right-handed vector currents. We follow the formalism in \cite{Doi:1985dx} and adapt it to our scenario. Considering the Lagrangian in Eq.~\eqref{eq:lagrangian} of the main text, $\vvbb$ decay occurs at second order of the perturbative expansion; namely, the matrix element is in general given by
\begin{eqnarray}
       \mathcal{M}^{2\nu}\equiv \langle e_1 e_2 \bar\nu_1 \bar\nu_2 f|S^{(2)}|i\rangle 
       &=& \frac{(-i)^2}{2} 
	\int d^4x d^4y \langle e_1 e_2 \bar\nu_1 \bar\nu_2 f| \mathcal{T}\left[\mathcal{L}(x)\mathcal{L}(y)\right]|i\rangle,
        \nn\\
        &=&  \mathcal{M}^{2\nu}_{\rm SM} + \mathcal{M}^{2\nu}_\epsilon + \dots
\end{eqnarray}
and it contains both the SM contribution and the exotic contribution proportional to $\epsilon_{XR}$. Further, $\mathcal{T}$ denotes the time-ordered product
\begin{align}
	\mathcal{T}[\mathcal{L}(x)\mathcal{L}(y)] =   
	\Theta(x_0-y_0)\mathcal{L}(x)\mathcal{L}(y)
	+\Theta(y_0-x_0)\mathcal{L}(y)\mathcal{L}(x),
\end{align}
and the initial and final states are composed of the decaying nucleus $|i\rangle$ and the final nucleus $|f\rangle$ together with the emitted electrons $e_{1,2}$ and antineutrinos $\bar\nu_{1,2}$. The integrations are over the space-time coordinates $x$ and $y$ of the two interactions involved.

We here concentrate on the case with one SM interaction and one exotic right-handed interaction. The matrix element can then be expressed as
\begin{align}
	\mathcal{M}^{2\nu}_{\epsilon} &= 
	(-i)^2 \left(\frac{1}{\sqrt{2}}\right)^2 
	\left(\frac{G_\beta}{\sqrt{2}}\right)^2 
	\epsilon_{XR} [1-\text{P}(e_1,e_2)] [1-\text{P}(\bar\nu_1,\bar\nu_2)] \nn\\
	&\times \int d^4x d^4y \left[\overline{\psi}(p_{e_1},x)\gamma_\mu(1+\gamma_5)\psi_{\bar\nu}^{c}(p_{\bar\nu_1},x)\right] 
	 \left[\overline{\psi}(p_{e_2},y)\gamma_\nu(1-\gamma_5)\psi^{c}(p_{\bar\nu_2},y)\right]  \nn\\
	 &\quad\times \bigg[ \Theta(x_0-y_0)\sum_n \langle f| J^\mu_X(x) | n\rangle\langle n|J^{\nu}_L(y)|i \rangle +
           \Theta(y_0-x_0) \sum_n \langle f| J^\nu_L(y) | n\rangle\langle n|J^{\mu}_X(x)|i \rangle \bigg] \nn\\
     &\quad\times
             \left[\overline{\psi}(p_{e_1},y)\gamma_\nu(1-\gamma_5)\psi^{c}(p_{\bar\nu_1},y)\right]
             \left[\overline{\psi}(p_{e_2},x)\gamma_\mu(1+\gamma_5)\psi^{c}(p_{\bar\nu_2},x)\right],          
\end{align}
where $G_\beta=G_F\cos\Theta_C$ and $\mathrm{P}(a,b)$ is the permutation operator interchanging the particles $a$ and $b$. Further, $\psi(p,x)$ stands for the electron or antineutrino wave function with four momentum $p=(E,\mathbf{p})$ and position $x=(x_0,\mathbf{x})$, $J^\mu(x)_X$ denotes the nuclear current with chirality $X$ and $|n\rangle$ is the intermediate nucleus state. For calculating the matrix element of the SM contribution, one would only need to replace in the above expression the right-handed projector $(1+\gamma_5)$ in the first lepton current by a left-handed one and follow the subsequent derivation in an analogous manner.

Writing the time dependence of the wave functions and currents explicitly allows performing the integration over time variables $x_0$ and $y_0$ with the result
\begin{align}
	\mathcal{M}^{2\nu}_{\epsilon} &=  2\pi \delta(E_{e_1}+E_{e_2}+E_{\bar\nu_1}+E_{\bar\nu_2}+E_f-E_i)~\epsilon_{XR}~\mathcal{R}^{2\nu}_{\epsilon} \nn \\
	&= 2\pi \delta(E_{e_1}+E_{e_2}+E_{\bar\nu_1}+E_{\bar\nu_2}+E_f-E_i) 
	\,i\left(\frac{1}{\sqrt{2}}\right)^2 \left(\frac{G_\beta}{\sqrt{2}}\right)^2 
	\epsilon_{XR} \nn \\
    &\times [1-\text{P}(e_1,e_2)] [1-\text{P}(\bar\nu_1,\bar\nu_2)] \nn\\
	&\times \int d^3\mathbf{x} d^3\mathbf{y} \left[\overline{\psi}(p_{e_1},\mathbf{x})\gamma_\mu(1+\gamma_5)\psi^{c}(p_{\bar\nu_1},\mathbf{x})\right]
        \left[\overline{\psi}(p_{e_2},\mathbf{y})\gamma_\nu(1-\gamma_5)\psi^{c}(p_{\bar\nu_2},\mathbf{y})\right] \nn \\
	&\quad\times \bigg[ \sum_n \frac{\langle f| J^\mu_X(0,\mathbf{x})|n\rangle\langle n|J^\nu_L(0,\mathbf{y})|i\rangle}{E_n - E_i + E_{e_2} + E_{\bar\nu_2}}
         + \sum_n\frac {\langle f| J^\nu_L(0,\mathbf{y})|n\rangle\langle n|J^\mu_X(0,\mathbf{x})|i\rangle}{E_n - E_i + E_{e_2} + E_{\bar\nu_2}} \bigg] \nn\\
      &\quad\times \left[\overline{\psi}(p_{e_1},\mathbf{y})\gamma_\nu(1-\gamma_5)\psi^{c}(p_{\bar\nu_1},\mathbf{y})\right]
         \left[\overline{\psi}(p_{e_2},\mathbf{x})\gamma_\mu(1+\gamma_5)\psi^{c}(p_{\bar\nu_2},\mathbf{x})\right].
\label{eq:totM}
\end{align}
Here, $E_x$ denotes the energy of particle $x$ or respective nucleus, and the delta function guaranteeing energy conservation and energy denominator appear as a result of the integration over the time components.

Now we employ two approximations. First, we take the non-relativistic expansion of the nuclear currents,
\begin{align}
J^\mu(0,\mathbf{x})_{L/R}=\sum_m\tau^+_m~[ g_V g^{\mu 0} \mp g_A g^{\mu k} \sigma_m^k]~\delta(\mathbf{x}-\mathbf{x}_m),
\end{align}
where we ignored the induced currents for their negligible contribution. Here, $g_V$ and $g_A$ are the vector and effective axial-vector coupling constants, respectively. Second, for the purpose of a factorization of nuclear matrix elements and phase space integral calculation we assume a standard approximation in which lepton wave functions are replaced with their values $\psi(p)=\psi(p,R)$ at the nuclear surface. For a $0^+\to 0^+$, ground state to ground state, transition we get
\begin{align}
  \label{eq:totM1ap}
  \mathcal{R}^{2\nu}_{\epsilon}  &= i \left(\frac{1}{\sqrt{2}}\right)^2 \left(\frac{G_\beta}{\sqrt{2}}\right)^2
        [1-\text{P}(e_1,e_2)] [1-\text{P}(\bar\nu_1,\bar\nu_2)] \nn \\
	&\times \left[~\overline{\psi}(p_{e_1})\gamma^\mu(1+\gamma_5)\psi^{c}(p_{\bar\nu_1})~
          \overline{\psi}(p_{e_2})\gamma^\nu(1-\gamma_5)\psi^{c}(p_{\bar\nu_2}) \right. \nn \\
	&~~+ \left. \overline{\psi}(p_{e_1})\gamma^\nu(1-\gamma_5)\psi^{c}(p_{\bar\nu_1})~
          \overline{\psi}(p_{e_2})\gamma^\mu(1+\gamma_5)\psi^{c}(p_{\bar\nu_2}) \right] \nn \\          
	&\times \bigg[g_{\mu 0} g_{\nu 0}~ g_V^2~\sum_n \frac{M_F(n)}{E_n - E_i + E_{e_2} + E_{\bar\nu_2}} \mp\frac{1}{3} g_{\mu k} g_{\nu k}~ g_A^2~
          \sum_n \frac{M_{GT}(n)}{E_n - E_i + E_{e_2} + E_{\bar\nu_2}} \bigg]
\end{align}
with
\begin{align}
	M_F(n) &=
	\langle 0^+_f | \sum_j\tau^+_j | 0^+_n\rangle
	\langle 0^+_n | \sum_k\tau^+_k | 0^+_i\rangle, \nn\\
	M_{GT}(n) &=
	\langle 0^+_f | \sum_j\tau^+_j \boldsymbol\sigma_j | 1^+_n\rangle\cdot
	\langle 1^+_n | \sum_k\tau^+_k \boldsymbol\sigma_k | 0^+_i\rangle.
\end{align}

The sign of the $g_A^2$-proportional part depends on the chirality $X$ of the quark current appearing in the exotic effective interaction $\epsilon_{XR}$ -- it is negative (positive) for a left-handed (right-handed) quark current. Further, we specify the angular momentum and parity of the nuclear states with $|0^+_i\rangle$, $|0^+_f\rangle$ denoting the $0^+$ ground states of the initial and final even-even nuclei, respectively. The intermediate nucleus states are denoted $|0^+_n\rangle$ ($|1^+_n\rangle$) for all possible levels $n$ with angular momentum and parity $J^\pi = 0^+$ ($J^\pi = 1^+$) with the corresponding energy $E_n$. The isospin-raising operators for a given nucleon $j$ is denoted as $\tau^+_j$, summed over all nucleons in the initial and final states. Likewise, $\vecs{\sigma}_j$ stands for the spin operator of nucleon $j$.

By writing out explicitly all terms in Eq.~\eqref{eq:totM1ap} we find
\begin{align}
  \mathcal{R}^{2\nu}_{\epsilon} &= i \left(\frac{G_{\beta}}{\sqrt{2}}\right)^2 \frac{1}{m_e} \nn\\
	&\times \bigg\{g_V^2 M^K_F \left[ \overline{\psi}(p_{e_1})\gamma_0(1+\gamma_5)\psi^{c}(p_{\bar\nu_1})
                                 ~\overline{\psi}(p_{e_2})\gamma_0(1-\gamma_5)\psi^{c}(p_{\bar\nu_2})\right. \nn \\
        &~~~~~~~~~~~~\left.+\,\overline{\psi}(p_{e_1})\gamma_0(1-\gamma_5)\psi^{c}(p_{\bar\nu_1})
                                  ~\overline{\psi}(p_{e_2})\gamma_0(1+\gamma_5)\psi^{c}(p_{\bar\nu_2})\right] \nn \\
	&~~\,-g_V^2 M^L_F \left[\overline{\psi}(p_{e_1})\gamma_0(1+\gamma_5)\psi^{c}(p_{\bar\nu_2})
                                 ~\overline{\psi}(p_{e_2})\gamma_0(1-\gamma_5)\psi^{c}(p_{\bar\nu_1})\right. \nn\\
        &~~~~~~~~~~~~\left.+\,\overline{\psi}(p_{e_1})\gamma_0(1-\gamma_5)\psi^{c}(p_{\bar\nu_2})
                                  ~\overline{\psi}(p_{e_2})\gamma_0(1+\gamma_5)\psi^{c}(p_{\bar\nu_1})\right] \nn\\
	&~~~\mp \frac{1}{3} g_A^2 M^K_{GT} \left[
	\overline{\psi}(p_{e_1})\gamma_k(1+\gamma_5)\psi^{c}(p_{\bar\nu_1})~
        \overline{\psi}(p_{e_2})\gamma_k(1-\gamma_5)\psi^{c}(p_{\bar\nu_2})\right.\nn\\
        &~~~~~~~~~~~~~~~~\,\left.+\, \overline{\psi}(p_{e_1})\gamma_k(1-\gamma_5)\psi^{c}(p_{\bar\nu_1})~
        \overline{\psi}(p_{e_2})\gamma_k(1+\gamma_5)\psi^{c}(p_{\bar\nu_2})\right] \nn \\
	&~~~\pm \frac{1}{3} g_A^2 M^L_{GT} \left[
	\overline{\psi}(p_{e_1})\gamma_k(1+\gamma_5)\psi^{c}(p_{\bar\nu_2})~
        \overline{\psi}(p_{e_2})\gamma_k(1-\gamma_5)\psi^{c}(p_{\bar\nu_1})\right. \nn \\
        &~~~~~~~~~~~~~~~~\,\left. +\, \overline{\psi}(p_{e_1})\gamma_k(1-\gamma_5)\psi^{c}(p_{\bar\nu_2})~
        \overline{\psi}(p_{e_2})\gamma_k(1+\gamma_5)\psi^{c}(p_{\bar\nu_1})\right]\bigg\}, 
\label{eq:totMperms}
\end{align}
where we define Fermi and Gamow-Teller nuclear matrix elements
\begin{align}
\label{fagtmatt}
	M^{K,L}_{F, GT} = m_e \sum_n M_{F, GT}(n)
	\frac{E_n - (E_i+E_f)/2}{[E_n - (E_i+E_f)/2]^2 - \varepsilon^2_{K,L}}.
\end{align}
Here, we conventionally put the electron mass $m_e$ to make the NMEs dimensionless.
The lepton energies enter in Eq.~\eqref{fagtmatt} through the terms
\begin{align}
  \varepsilon_K = 
  \frac{1}{2}\left(E_{e_2}+E_{\bar\nu_2}-E_{e_1}-E_{\bar\nu_1}\right),\qquad
  \varepsilon_L = 
  \frac{1}{2}\left(E_{e_1}+E_{\bar\nu_2}-E_{e_2}-E_{\bar\nu_1}\right),
\end{align}  
which range between $-Q/2 \leq \varepsilon_{K,L} \leq Q/2$. For $2\nu\beta\beta$ decay with energetically forbidden transitions to the intermediate states, $E_n - E_i > - m_e$, the quantity $E_n -(E_i+E_f)/2 = Q/2 + m_e + (E_n-E_i)$ is always larger than $Q/2$.

We first focus on the leptonic part of the total matrix element. Employing the equivalence
\begin{align}
	  \overline{\psi}(p_{e_1})\gamma_{\mu}(1-\gamma_5)\psi^{c}(p_{\bar\nu_1}) 
	= \overline{\psi}(p_{\bar\nu_1})\gamma_{\mu}(1+\gamma_5)\psi^c(p_{e_1}),
\label{eq:transpose}
\end{align}
and the Fierz transformation
\begin{align}
	 &\overline{\psi}(p_{e_1})\gamma_{\mu}(1+\gamma_5)\psi^{c}(p_{\bar\nu_1})~   
	  \overline{\psi}(p_{\bar\nu_2})\gamma_{\nu}(1+\gamma_5)\psi^c(p_{e_2}) \nn\\
	  &= \frac{1}{2}\overline{\psi}(p_{e_1})\gamma_\sigma(1+\gamma_5) \psi^c(p_{e_2})~
          \overline{\psi}(p_{\bar\nu_2})\gamma_{\mu}\gamma^\sigma\gamma_{\nu}(1+\gamma_5)\psi^{c}(p_{\bar\nu_1}),
\end{align}
to all four permuted terms in Eq.~\eqref{eq:totMperms}, and using the identity $\gamma^\alpha\gamma_\mu\gamma_\alpha = -2 \gamma_\mu$
one obtains the reaction matrix element in the following form
\begin{align}
  \mathcal{R}^{2\nu}_{\epsilon} &= i \frac{G_\beta^2}{4} \frac{1}{m_e} \nn \\
    &\times \bigg\{ g_V^2 (M^K_F - M^L_F) \overline{\psi}(p_{e_1})\gamma^{\sigma}\psi^{c}(p_{e_2}) 
	 ~\overline{\psi}(p_{\bar\nu_1})\gamma_{0}\gamma_{\sigma}\gamma^{0} \psi_{\bar\nu}^{c}(p_{\bar\nu_2}) \nn\\
	&~~\,-g_V^2(M^K_F + M^L_F) \overline{\psi}(p_{e_1})\gamma^{\sigma}\gamma_5 \psi_{\bar\nu}(p_{e_2})
        ~\overline{\psi}(p_{\bar\nu_1})\gamma_{0}\gamma_{\sigma}\gamma_{0}\gamma_5 \psi^{c}(p_{\bar\nu_2}) \nn\\
	&~~\, \pm \frac{1}{3} \bigg[ 2 g_A^2 (M^K_{GT} - M^L_{GT})  
	\big[\overline{\psi}(p_{e_1})\gamma^{\sigma}\psi^{c}(p_{e_2}) ~ \overline{\psi}(p_{\bar\nu_1})\gamma_{\sigma}\psi^{c}(p_{\bar\nu_2}) \nn\\
	&\hspace{4.25cm}+\overline{\psi}(p_{e_1})\gamma^{\sigma}\psi^{c}(p_{e_2})   
	     ~\overline{\psi}(p_{\bar\nu_1})\gamma_{0}\gamma_{\sigma}\gamma_{0}\psi^{c}(p_{\bar\nu_2}) \big] \nn\\
&~~~~~~\,-2 g_A^2 (M^K_{GT} + M^L_{GT}) \big[ \overline{\psi}(p_{e_1})\gamma^{\sigma}\gamma_{5}\psi^{c}(p_{e_2})
                                  ~\overline{\psi}(p_{\bar\nu_1})\gamma_{\sigma}\gamma_{5}\psi^{c}(p_{\bar\nu_2}) \nn\\
&\hspace{4.25cm}+\overline{\psi}(p_{e_1})\gamma^{\sigma}\gamma_{5}\psi^{c}(p_{e_2})
~\overline{\psi}(p_{\bar\nu_1})\gamma_{0}\gamma_{\sigma}\gamma^{0}\gamma_{5}\psi_{\bar\nu}^{c}(p_{\bar\nu_2}) \big]\bigg]\bigg\}.
\label{eq:s2permsFT}
\end{align}

In the following, we consider the $S_{1/2}$ spherical wave approximation for the outgoing electrons, i.e.
\begin{align}
\label{eq:radialwf}
	\psi_s({p}_e) = \begin{pmatrix} 
		g_{-1}(E_e)\chi_s \\ 
		 f_{+1}(E_e)\left(\vecs{\sigma}\cdot\hat{\vecl{p}}_e\right)\chi_s 
	\end{pmatrix},
\end{align}
where $\chi_s$ is a two-component spinor, $\hat{\vecl{p}}_e=\vecl{p}_e/|\vecl{p}_e|$ stands for the direction of the electron momentum and $g_{-1}(E_e)$ and $f_{+1}(E_e)$ are the radial electron wave functions depending on the electron energy $E_e$ and evaluated at the nucleus' surface, i.e. at distance $R$ from the centre of the nucleus. On the other hand, as neutrinos do not feel the electromagnetic potential of the nucleus, they are considered to be plane waves in long-wave approximation,
\begin{align}
  \psi(p_\nu) =
  \sqrt{\frac{{E_\nu + m_\nu}}{{2 E_\nu}}}
  \left(\begin{array}{c} 
		      \chi_s \\ 
		      \frac{\left(\vecs{\sigma}\cdot\hat{\mathbf{p}}_\nu\right)}{E_\nu + m_\nu} \chi_s
	\end{array}\right).
\label{eq:vplane} 
\end{align}

We now take the square of the absolute value of the matrix element in Eq.~\eqref{eq:s2permsFT}, using the wave functions in Eqs.~\eqref{eq:radialwf} and \eqref{eq:vplane}, and sum over the spins. After evaluating those and keeping only the terms which do not vanish when integrating over neutrino momenta, we are left with a somewhat lengthy expression,
\begin{align}
\sum_\text{spins}|\mathcal{R}^{2\nu}_{\epsilon}|^2 &=\frac{G_\beta^2}{8}\bigg\{ g_V^4 (M_F^K-M_F^L)^2 \big[ 4 E_{\nu_1} E_{\nu_2} \tilde{E}_{e_1} \tilde{E}_{e_2} + 2 \tilde{m}_{e}^2 E_{\nu_1} E_{\nu_2} \nn \\
&~~~~~\phantom{\times \bigg\{ g_V^4 (M_F^K-M_F^L)^2\big[} + 2 m_{\nu}^2 \tilde{E}_{e_1} \tilde{E}_{e_2} + 4 m_{\nu}^2 \tilde{m}_{e}^2 - 2 m_{\nu}^2 (\tilde{\vecl{p}}_{e_1} \cdot \tilde{\vecl{p}}_{e_2}) \big] \nn \\
&~~~~~\hspace{0.35cm} +g_V^4 (M_F^K+M_F^L)^2 \big[ 4 E_{\nu_1} E_{\nu_2} \tilde{E}_{e_1} \tilde{E}_{e_2} - 2 \tilde{m}_{e}^2 E_{\nu_1} E_{\nu_2} \nn \\
&~~~~~\phantom{\times \bigg\{ g_V^4 (M_F^K+M_F^L)^2 \big[}- 2 m_{\nu}^2 \tilde{E}_{e_1} \tilde{E}_{e_2} + 4 m_{\nu}^2 \tilde{m}_{e}^2 + 2 m_{\nu}^2 (\tilde{\vecl{p}}_{e_1} \cdot \tilde{\vecl{p}}_{e_2}) \big] \nn \\
&~~~~~\hspace{0.35cm} \pm g_V^2 g_A^2 (M_F^K-M_F^L) (M_{GT}^K-M_{GT}^L) \big[ - 12 \tilde{m}_{e}^2 E_{\nu_1} E_{\nu_2} - 12 m_{\nu}^2 \tilde{E}_{e_1} \tilde{E}_{e_2} \nn \\
&~~~~~\phantom{\times \bigg\{ g_V^4 (M_F^K+M_F^L)^2 \big[}+8 E_{\nu_1} E_{\nu_2} (\tilde{\vecl{p}}_{e_1} \cdot \tilde{\vecl{p}}_{e_2}) - 4 m_{\nu}^2 (\tilde{\vecl{p}}_{e_1} \cdot \tilde{\vecl{p}}_{e_2}) \big] \nn \\
&~~~~~\hspace{0.35cm} \pm g_V^2 g_A^2 (M_F^K+M_F^L) (M_{GT}^K+M_{GT}^L) \big[ 12 \tilde{m}_{e}^2 E_{\nu_1} E_{\nu_2} + 12 m_{\nu}^2 \tilde{E}_{e_1} \tilde{E}_{e_2} \nn \\
&~~~~~\phantom{\times \bigg\{ g_V^4 (M_F^K+M_F^L)^2 \big[}+8 E_{\nu_1} E_{\nu_2} (\tilde{\vecl{p}}_{e_1} \cdot \tilde{\vecl{p}}_{e_2}) + 4 m_{\nu}^2 (\tilde{\vecl{p}}_{e_1} \cdot \tilde{\vecl{p}}_{e_2}) \big] \nn \\
&~~~~~\hspace{0.35cm}+g_A^4 (M_{GT}^K-M_{GT}^L)^2 \big[ 12 E_{\nu_1} E_{\nu_2} \tilde{E}_{e_1} \tilde{E}_{e_2} - 6 \tilde{m}_{e}^2 E_{\nu_1} E_{\nu_2} - 6 m_{\nu}^2 \tilde{E}_{e_1} \tilde{E}_{e_2} \nn \\
&~~~~~\phantom{\times \bigg\{ g_V^4 (M_F^K-M_F^L)^2\big[} + 12 m_{\nu}^2 \tilde{m}_{e}^2 + 8 E_{\nu_1} E_{\nu_2} (\tilde{\vecl{p}}_{e_1} \cdot \tilde{\vecl{p}}_{e_2}) - 10 m_{\nu}^2 (\tilde{\vecl{p}}_{e_1} \cdot \tilde{\vecl{p}}_{e_2}) \big] \nn \\
&~~~~~\hspace{0.35cm}+g_A^4 (M_{GT}^K+M_{GT}^L)^2 \big[ 12 E_{\nu_1} E_{\nu_2} \tilde{E}_{e_1} \tilde{E}_{e_2} + 6 \tilde{m}_{e}^2 E_{\nu_1} E_{\nu_2} + 6 m_{\nu}^2 \tilde{E}_{e_1} \tilde{E}_{e_2} \nn \\
&~~~~~\phantom{\times \bigg\{ g_V^4 (M_F^K-M_F^L)^2\big[} + 12 m_{\nu}^2 \tilde{m}_{e}^2 + 8 E_{\nu_1} E_{\nu_2} (\tilde{\vecl{p}}_{e_1} \cdot \tilde{\vecl{p}}_{e_2}) + 10 m_{\nu}^2 (\tilde{\vecl{p}}_{e_1} \cdot \tilde{\vecl{p}}_{e_2}) \big]\bigg\}.
\label{eq:longME}
\end{align}
Here, the terms proportional to $m_\nu^2$ can be safely omitted for light active neutrinos with $m_\nu \lesssim 0.1$~eV. The dependence on the electron radial wave functions $f_1(E_{e_i})$,  $g_{-1}(E_{e_i})$ is contained in the terms
\begin{align}
	\tilde{E}_i 
	&= E_{e_i} [g_{-1}^2(E_{e_i})+f_1^2(E_{e_i})] 
	 \simeq E_{e_i} F_0(Z_f, E_{e_i}),  \nn\\
	\tilde{\vecl{p}}_{e_i} 
	&= \vecl{p}_{e_i} \frac{2E_{e_i}}{|\vecl{p}_{e_i}|} f_1(E_{e_i})g_{-1}(E_{e_i})
	 \simeq  \vecl{p}_{e_i} F_0(Z_f, E_{e_i}), \\
	\tilde{m}_e
	&= E_e [g_{-1}^2(E_{e})-f_1^2(E_{e})]
	 \simeq m_e F_0(Z_f, E_{e_i}). \nn
\end{align}
In our numerical calculations we employ the above shown approximations using the relativistic Fermi function $F_0(Z_f, E_e)$ for each electron of energy $E_{e_i}$ and spatial momentum $p_{e_i}$ of the form \cite{Doi:1985dx}
\begin{align}
\label{eq:fermi}
	F_0(Z_f, E_e) 
	= \left(\frac{2}{\Gamma(1+2\gamma_0)}\right)^2 
	  (2 p_e R)^{2(\gamma_0-1)}e^{\pi y}|\Gamma(\gamma_0+iy)|^2,
\end{align}
with $\gamma_0 = \sqrt{1-(Z_f \alpha)^2}$ and $y = \alpha Z_f E_e / p_e$ where $Z_f = Z+2$ is the charge number of the final nucleus, $\alpha$ denotes the fine structure constant, $R$ is the nuclear radius and $\Gamma(x)$ stands for the Gamma function. The results obtained using this approximation do not deviate from the more accurate radial electron wave functions coming from the numerical solution of the Dirac equation by more than $\sim 10-15\%$ and the change in the angular correlation between the electron energies is negligible \cite{Kotila:2012zza}.
 
Equation~\eqref{eq:longME} can now be mapped to the coefficients $\mathcal{A}^{2\nu}_{\epsilon}$ and $\mathcal{B}^{2\nu}_{\epsilon}$ entering the differential decay rate Eq.~\eqref{eq:diffrate} of the process. For the terms independent of the scalar product of the spatial electron momenta this gives
\begin{align}
	{\cal A}^{2\nu}_{\epsilon} &=
  	4 \bigg\{ \left[ g_V^4 (M_F^K - M_F^L)^2 
  	+ \frac{1}{3}g_A^4 (M_{GT}^K - M_{GT}^L)^2 \right] \nn\\
  	&\quad\,\,\,+\left[g_V^4 (M_F^K + M_F^L)^2 + \frac{1}{3}g_A^4 (M_{GT}^K 
  	 + M_{GT}^L)^2 \right] \bigg\} \nn\\
	&\quad\,\,\, \times [g_{-1}^2(E_{e_1})+f_1^2(E_{e_1})][g_{-1}^2(E_{e_2})+f_1^2(E_{e_2})] \nn\\
	&+2\bigg\{ \left[ g_V^4 (M_F^K - M_F^L)^2 
	 -\frac{1}{3}g_A^4 (M_{GT}^K - M_{GT}^L)^2 \right] \nn\\
	&\quad\,\,\,-\left[ g_V^4 (M_F^K + M_F^L)^2 - \frac{1}{3}g_A^4 (M_{GT}^K 
	 + M_{GT}^L)^2 \right] \nn\\ 
	&\quad\,\,\,+ 2 g_V^2g_A^2 \left[(M_F^K - M_F^L)(M_{GT}^K - M_{GT}^L) 
	 + (M_F^K + M_F^L)(M_{GT}^K + M_{GT}^L) \right] \bigg\} \nn\\
	&\quad\,\,\, \times [g_{-1}^2(E_{e_1})-f_1^2(E_{e_1})][g_{-1}^2(E_{e_2})-f_1^2(E_{e_2})].
\label{eq:Aright}
\end{align}
Here, the dependence on the electron radial wave functions has been made explicit. Likewise, the terms proportional to $\hat{\vecl{p}}_{e_1}\cdot \hat{\vecl{p}}_{e_2} = \cos\theta$ combine to give
\begin{align}
	{\cal B}^{2\nu}_{\epsilon} &=
	\bigg\{\frac{8}{9} g_A^4 \left[ (M_{GT}^K - M_{GT}^L)^2 + (M_{GT}^K 
	 + M_{GT}^L)^2 \right]  \nn \\
   &~~\, - \frac{8}{3} g_V^2g_A^2 \left[(M_F^K - M_F^L)(M_{GT}^K 
	 - M_{GT}^L) + (M_F^K + M_F^L)(M_{GT}^K + M_{GT}^L) \right]
	 \bigg\} \nn \\
	 &\quad\times 4f_1(E_{e_1})f_1(E_{e_2})g_{-1}(E_{e_1})g_{-1}(E_{e_2}),
\label{eq:Bright}
\end{align}
The above results may be further simplified in well-motivated approximations.

\paragraph*{Isospin Invariance:} In the approximate case of isospin conservation one can set $M_F^K = M_F^L = 0$ (i.e. the double Fermi nuclear matrix elements vanish). Equations~\eqref{eq:Aright} and \eqref{eq:Bright} then simplify to, respectively,
\begin{align}
	{\cal A}^{2\nu}_{\epsilon} &\simeq
	\frac{4}{3} g_A^4 \left[ (M_{GT}^K + M_{GT}^L)^2 + (M_{GT}^K - M_{GT}^L)^2 \right] [g_{-1}^2(E_{e_1})+f_1^2(E_{e_1})][g_{-1}^2(E_{e_2})+f_1^2(E_{e_2})] \nn \\
	&+\, \frac{2}{3} g_A^4 \left[ (M_{GT}^K + M_{GT}^L)^2 - (M_{GT}^K - M_{GT}^L)^2 \right] 
%\nn \\ &~~~~~~\,\times 
[g_{-1}^2(E_{e_1})-f_1^2(E_{e_1})][g_{-1}^2(E_{e_2})-f_1^2(E_{e_2})],
\end{align}
and
\begin{align}
	{\cal B}^{2\nu}_{\epsilon} &\simeq
	\frac{8}{9} g_A^4 \left[ (M_{GT}^K + M_{GT}^L)^2 + (M_{GT}^K - M_{GT}^L)^2 \right] 4f_1(E_{e_1})f_1(E_{e_2})g_{-1}(E_{e_1})g_{-1}(E_{e_2}).
\end{align}

\paragraph*{Neglecting lepton energies in NMEs:} In the case that $\varepsilon_{K,L}$ are neglected in the energy denominators of NMEs, the nuclear and leptonic parts can be treated separately and the result simplifies to
\begin{align}\label{nucmata}
	{\cal A}^{2\nu}_{\epsilon} &\simeq
	\frac{8}{3} g_A^4 M_{GT} \big\{ 3 [f_1^2(E_{e_1})f_1^2(E_{e_2})+g_{-1}^2(E_{e_1})g_{-1}^2(E_{e_2})] \nn \\
	&\hspace{2.1cm}+ [f_1^2(E_{e_1})g_{-1}^2(E_{e_2})+g_{-1}^2(E_{e_1})f_1^2(E_{e_2})] \big\}, \nn \\
	{\cal B}^{2\nu}_{\epsilon} &\simeq
	\frac{32}{9} g_A^4 M_{GT} ~4f_1(E_{e_1})f_1(E_{e_2})g_{-1}(E_{e_1})g_{-1}(E_{e_2}),
\end{align}
with the Gamow-Teller nuclear matrix element now given by
\begin{align}
	M_{GT} &= m_e \sum_n 
	\frac{\langle 0^+_f | \sum_{m}\tau^+_m 
		\sigma_m | 1^+_{n}\rangle
	\langle 1^+_n | \sum_{m}\tau^+_m \sigma_m | 0^+_{i} \rangle}
	{E_n - {(E_i+E_f)}/{2}}.
\end{align}

\paragraph*{Higher order corrections in lepton energies:} A more accurate expression can be obtained by Taylor expanding the nuclear matrix elements in the small parameters $\varepsilon_{K,L}$~\cite{Simkovic:2018rdz}. Taking the series up to the fourth power in $\varepsilon_{K,L}$ we get
\begin{align}
\label{eq:expampA}
	{\cal A}^{2\nu}_{\epsilon} &\simeq
	\frac{16}{3} g_A^4 \bigg[ (M_{GT-1})^2 + (\varepsilon_K^2+\varepsilon_L^2)M_{GT-1}M_{GT-3} \nn \\
	&\hspace{0.4cm}~~~~~~ + (\varepsilon_K^4+\varepsilon_L^4)\left(M_{GT-1}M_{GT-5}+\frac{1}{2}(M_{GT-3})^2\right) \bigg] \nn \\
	&\hspace{0.4cm}~~~~~~ \times [g_{-1}^2(E_{e_1})+f_1^2(E_{e_1})][g_{-1}^2(E_{e_2})+f_1^2(E_{e_2})] \nn \\
	&~~ + \frac{8}{3} g_A^4 \big[ (M_{GT-1})^2 + (\varepsilon_K^2+\varepsilon_L^2)M_{GT-1}M_{GT-3} \nn \\
	&\hspace{0.4cm}~~~~~~ + \varepsilon_K^2\varepsilon_L^2(M_{GT-3})^2 + (\varepsilon_K^4+\varepsilon_L^4) M_{GT-1}M_{GT-5} \big] \nn \\
	&\hspace{0.4cm}~~~~~~ \times [g_{-1}^2(E_{e_1})-f_1^2(E_{e_1})][g_{-1}^2(E_{e_2})-f_1^2(E_{e_2})]
\end{align}
and
\begin{align}
\label{eq:expampB}
	{\cal B}^{2\nu}_{\epsilon} &\simeq
	\frac{32}{9} g_A^4 \bigg[ (M_{GT-1})^2 + (\varepsilon_K^2+\varepsilon_L^2)M_{GT-1}M_{GT-3} \nn \\
	&\hspace{0.4cm} ~~~~~~ + (\varepsilon_K^4+\varepsilon_L^4)\left(M_{GT-1}M_{GT-5}+\frac{1}{2}(M_{GT-3})^2\right) \bigg] \nn \\
	&\hspace{0.4cm} ~~~~~~ \times 4f_1(E_{e_1})f_1(E_{e_2})g_{-1}(E_{e_1})g_{-1}(E_{e_2}).
\end{align}
Here, the introduced NMEs are defined as
\begin{align}
	M_{GT-1} &= M_{GT}, \label{eq:nme1} \\
	M_{GT-3} &= m_e^3 \sum_n \frac{4M_{GT}(n)}{(E_n-(E_i+E_f)/2)^3}, \\
	M_{GT-5} &= m_e^5 \sum_n \frac{16M_{GT}(n)}{(E_n-(E_i+E_f)/2)^5}.
	\label{eq:nme5}
\end{align}

\subsection*{A.2 Standard Model contribution}
The standard contribution to $\vvbb$ decay can be calculated likewise in our formalism. It arises from the first term in the Lagrangian in Eq.~\eqref{eq:lagrangian} of the main text, with the calculation proceeding analogously, essentially replacing $\epsilon_{XR} \to 1+\epsilon_{LL}$ and using $V-A$ currents throughout. The corresponding coefficients in Eq.~\eqref{eq:diffrate} are
\begin{align}
\label{eq:A-SM}
	{\cal A}^{2\nu}_{\rm SM} &=
	\bigg\{\frac{1}{4}\left[g_V^2 \left(M^K_{F} + M^L_{F}\right)
	- g_A^2 \left(M^K_{GT} + M^L_{GT}\right)  \right]^2 \nn\\
	&~~\, + \frac{3}{4}\left[ g_V^2 \left(M^K_{F} - M^L_{F}\right)
	+ \frac{1}{3} g_A^2 \left(M^K_{GT} - M^L_{GT}\right) \right]^2 \bigg\} \nn\\
	&~~\, \times [g_{-1}^2(E_{e_1})+f_1^2(E_{e_1})][g_{-1}^2(E_{e_2})+f_1^2(E_{e_2})],
\end{align}
and
\begin{align}
\label{eq:B-SM}
	{\cal B}^{2\nu}_{\rm SM} &=
	\bigg\{ \frac{1}{4}\left[ g_V^2 \left(M^K_{F} + M^L_{F}\right)
	- g_A^2 \left(M^K_{GT} + M^L_{GT}\right)  \right]^2 \nn \\
	&~~\, - \frac{1}{4}\left[ g_V^2 \left(M^K_{F} - M^L_{F}\right) + \frac{1}{3} g_A^4 \left(M^K_{GT} - M^L_{GT}\right) \right]^2 \bigg\} \nn \\
	&~~\, \times 4f_1(E_{e_1})f_1(E_{e_2})g_{-1}(E_{e_1})g_{-1}(E_{e_2}).
\end{align}
These results match with the literature~\cite{Haxton:1985am,Simkovic:2018rdz}.

\subsection*{A.3 Contribution from Standard Model -- Exotic interference}
Finally, the interference between the exotic and SM contributions enters the total $\vvbb$ rate; however, as mentioned, based on helicity considerations it is expected to be suppressed by $m_\nu / Q$ and thus be negligible for the emission of light eV-scale neutrinos. For completeness, we have calculated the corresponding term ${\cal A}^{2\nu}_{\epsilon \text{SM}}$ to verify the overall suppression by the light neutrino mass,
\begin{align}
{\cal A}^{2\nu}_{\epsilon{\rm SM}} =&
	\left[ g_V^2 \left(M^K_{F} + M^L_{F}\right) - g_A^2 \left(M^K_{GT} + M^L_{GT}\right)\right] \left[ g_V^2 \left(M^K_{F} - M^L_{F}\right) + 3 g_A^2 \left(M^K_{GT} - M^L_{GT}\right)\right] \nn \\
	&\times \frac{m_{\nu}\left(E_{\nu_2}-E_{\nu_1}\right)}{E_{\nu_1} E_{\nu_2}} [f_1^2(E_{e_1}) g_{-1}^2(E_{e_2}) - f_1^2(E_{e_2}) g_{-1}^2(E_{e_1})] \nn \\
	-&\left[ g_V^2 \left(M^K_{F} + M^L_{F}\right) - g_A^2 \left(M^K_{GT} + M^L_{GT}\right)\right] \left[ g_V^2 \left(M^K_{F} + M^L_{F}\right) + 3 g_A^2 \left(M^K_{GT} + M^L_{GT}\right)\right] \nn \\
	&\times \frac{m_{\nu}\left(E_{\nu_1}+E_{\nu_2}\right)}{E_{\nu_1} E_{\nu_2}}
	[f_1^2(E_{e_1}) f_{-1}^2(E_{e_2}) - g_1^2(E_{e_1}) g_{-1}^2(E_{e_2})] \nn \\
	+& \left[ g_V^2 \left(M^K_{F} - M^L_{F}\right) + \frac{1}{3} g_A^2 \left(M^K_{GT} - M^L_{GT}\right)\right] \left[ 3 g_V^2 \left(M^K_{F} + M^L_{F}\right) + 9 g_A^2 \left(M^K_{GT} + M^L_{GT}\right)\right] \nn \\
	&\times \frac{m_{\nu}\left(E_{\nu_2}-E_{\nu_1}\right)}{E_{\nu_1} E_{\nu_2}} [f_1^2(E_{e_1}) g_{-1}^2(E_{e_2}) - f_1^2(E_{e_2}) g_{-1}^2(E_{e_1})] \nn \\
	-& \left[ g_V^2 \left(M^K_{F} - M^L_{F}\right) + \frac{1}{3} g_A^2 \left(M^K_{GT} - M^L_{GT}\right)\right] \left[ 3 g_V^2 \left(M^K_{F} - M^L_{F}\right) + 9 g_A^2 \left(M^K_{GT} - M^L_{GT}\right)\right] \nn \\
	&\times \frac{m_{\nu}\left(E_{\nu_1}+E_{\nu_2}\right)}{E_{\nu_1} E_{\nu_2}}
	[f_1^2(E_{e_1}) f_{-1}^2(E_{e_2}) - g_1^2(E_{e_1}) g_{-1}^2(E_{e_2})].
\label{eq:Ainterference}
\end{align}
Moreover, the coefficient $\mathcal{B}^{2\nu}_{\epsilon{\rm SM}}$ determining the angular correlation of the outgoing electrons is in this case identically zero, ${\cal B}^{2\nu}_{\epsilon{\rm SM}} = 0$. In our numerical analysis we safely ignore the interference term.

\subsection*{A.4 Decay distributions and total rate}
The fully differential decay rate with respect to the (in principle) observable electron energies $E_{e_1}$, $E_{e_2}$ and the angle $\theta$ between their momenta is given by Eq.~\eqref{eq:diffrate}. The quantities $\mathcal{A}^{2\nu}$ and $\mathcal{B}^{2\nu}$ are calculated as discussed above, i.e. through Eqs.~\eqref{eq:A-SM} and \eqref{eq:B-SM} for the SM contribution and most importantly Eqs.~\eqref{eq:expampA} and \eqref{eq:expampB} for the exotic contribution quadratic in $\epsilon_{XR}$. In our numerical calculations we use the following physical constants: $G_\beta = 1.1363 \times 10^{-11}$~GeV$^{-2}$, $\alpha = 1/137$, $m_e = 0.511$~MeV, $m_p = 938$~MeV, $R = 1.2A^{1/3}$~fm (nucleon number $A=100$ for Molybdenum), $Q({}^{100}\mathrm{Mo})=3.03$~MeV, $g_V = 1$. For the axial coupling we take the value $g_A = 1$, as quenching of the usual value $g_A^{\rm nucleon}=1.269$ for a free neutron is expected in the nucleus~\cite{Gysbers:2019uyb}. In addition, we use the nuclear matrix elements for the $\vvbb$ decay of $^{100}$Mo from Ref.~\cite{Simkovic:2018rdz} given in Tab.~\ref{tab:nmes}.

\begin{figure}[t!]
	\centering
	\includegraphics[width=0.49\textwidth]{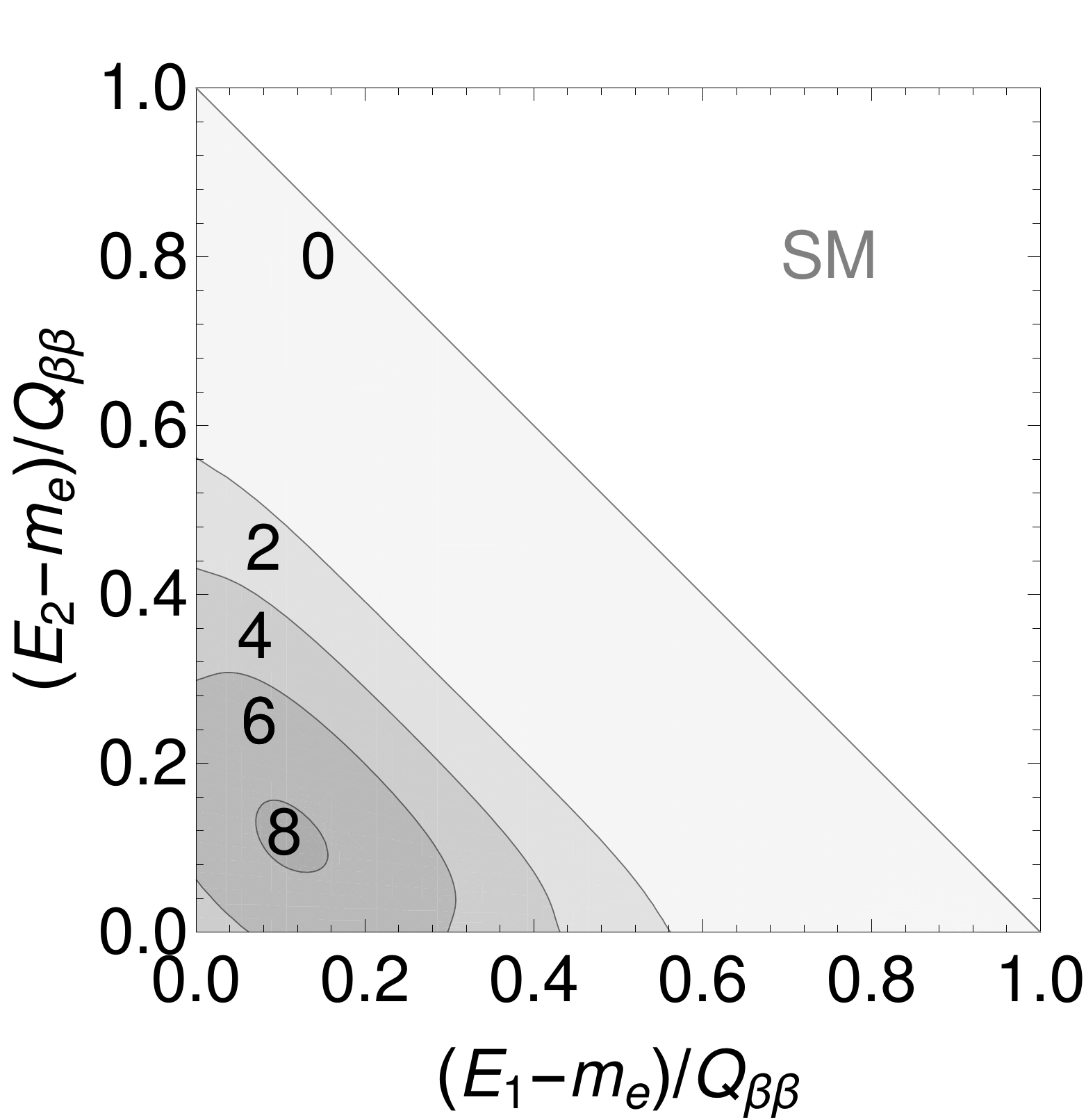}
	\hfill
	\includegraphics[width=0.49\textwidth]{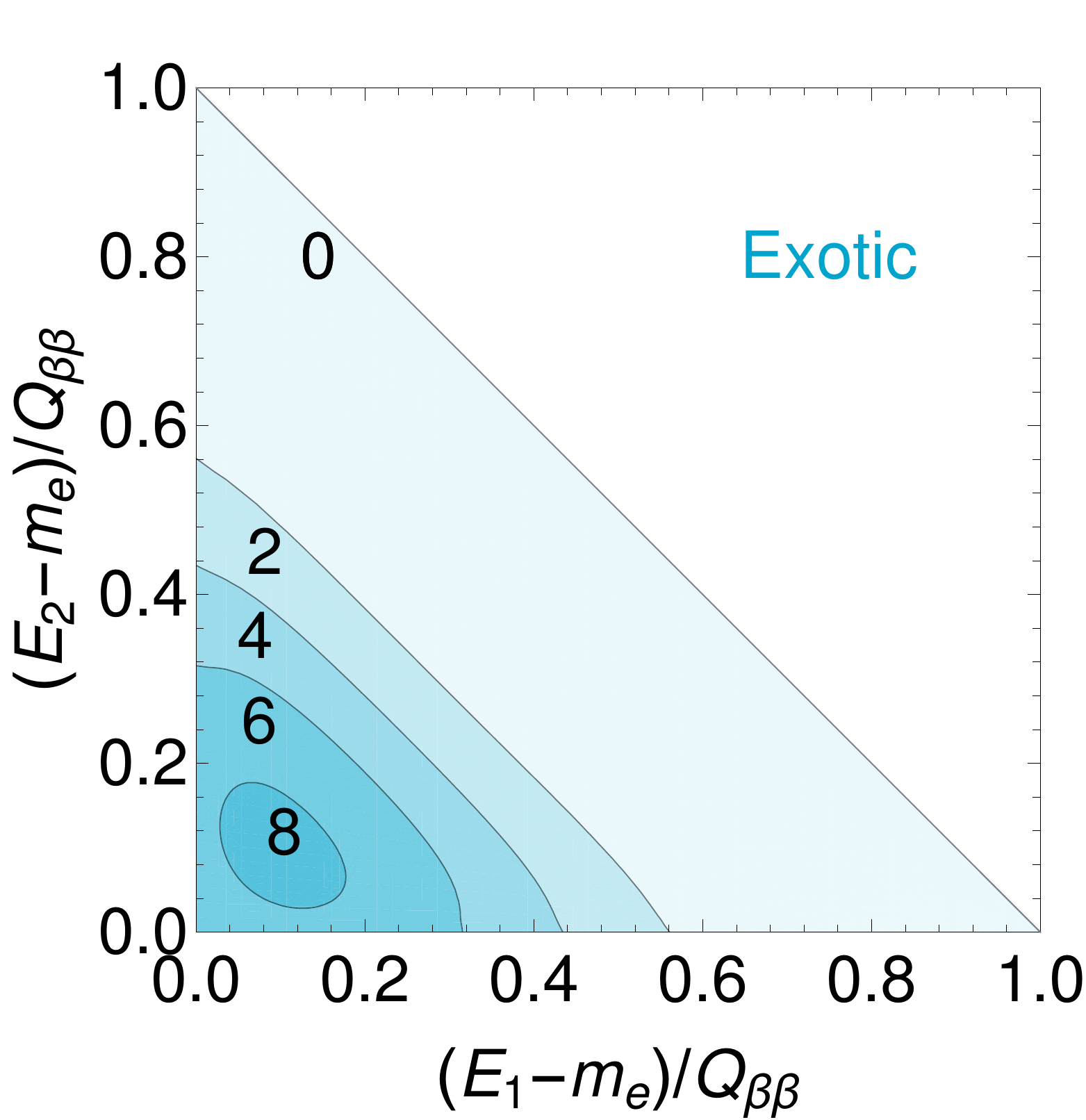}
	\caption{Normalized double energy distributions as functions of electron energies for SM $\vvbb$ decay (left) and for the exotic scenario incorporating a right-handed lepton current (right). Both plots are for $^{100}$Mo and the energies are normalized to the $Q$ value.}
	\label{fig:enDistros}
\end{figure}
\begin{table}[!b]
\centering 
\renewcommand{\arraystretch}{1.1}
\setlength\tabcolsep{0.2cm}
\begin{tabular}{rlllcccc}\hline\hline
Isotope & $M^{2\nu}_{GT-1}$ & $M^{2\nu}_{GT-3}$ & $M^{2\nu}_{GT-5}$ \\ \hline
$^{76}$Ge  & $0.111$  & $0.0133$  & $0.00263$ \\
$^{82}$Se  & $0.0795$ & $0.0129$  & $0.00355$ \\
$^{100}$Mo & $0.184^{+0.108}_{-0.072}$  & $0.0876^{+0.0354}_{-0.0243}$  & $0.0322^{+0.0131}_{-0.0089}$  \\
$^{136}$Xe & $0.0170$ & $0.00526$ & $0.00169$ \\
\hline\hline
\end{tabular}
\caption{Nuclear matrix elements used in our analysis (see Eqs.~\eqref{eq:nme1} to \eqref{eq:nme5}), calculated within the pn-QRPA with partial isospin restoration \cite{Simkovic:2018rdz} and using the effective axial coupling $g_A = 1.000$. For $^{100}$Mo we also show the uncertainty of the nuclear matrix elements when varying $g_A$ in the range $0.8 \leq g_A \leq 1.269$.
\label{tab:nmes}}
\end{table}

We now have all the ingredients to calculate the various decay distributions and the total decay rate potentially observable in double beta decay experiments.

\begin{figure}
	\centering
	\begin{minipage}{0.47\textwidth}
	\leavevmode
    \vbox{% baseline is at the bottom
    \offinterlineskip % disable line skip
    \hbox{\includegraphics[width=1\textwidth]{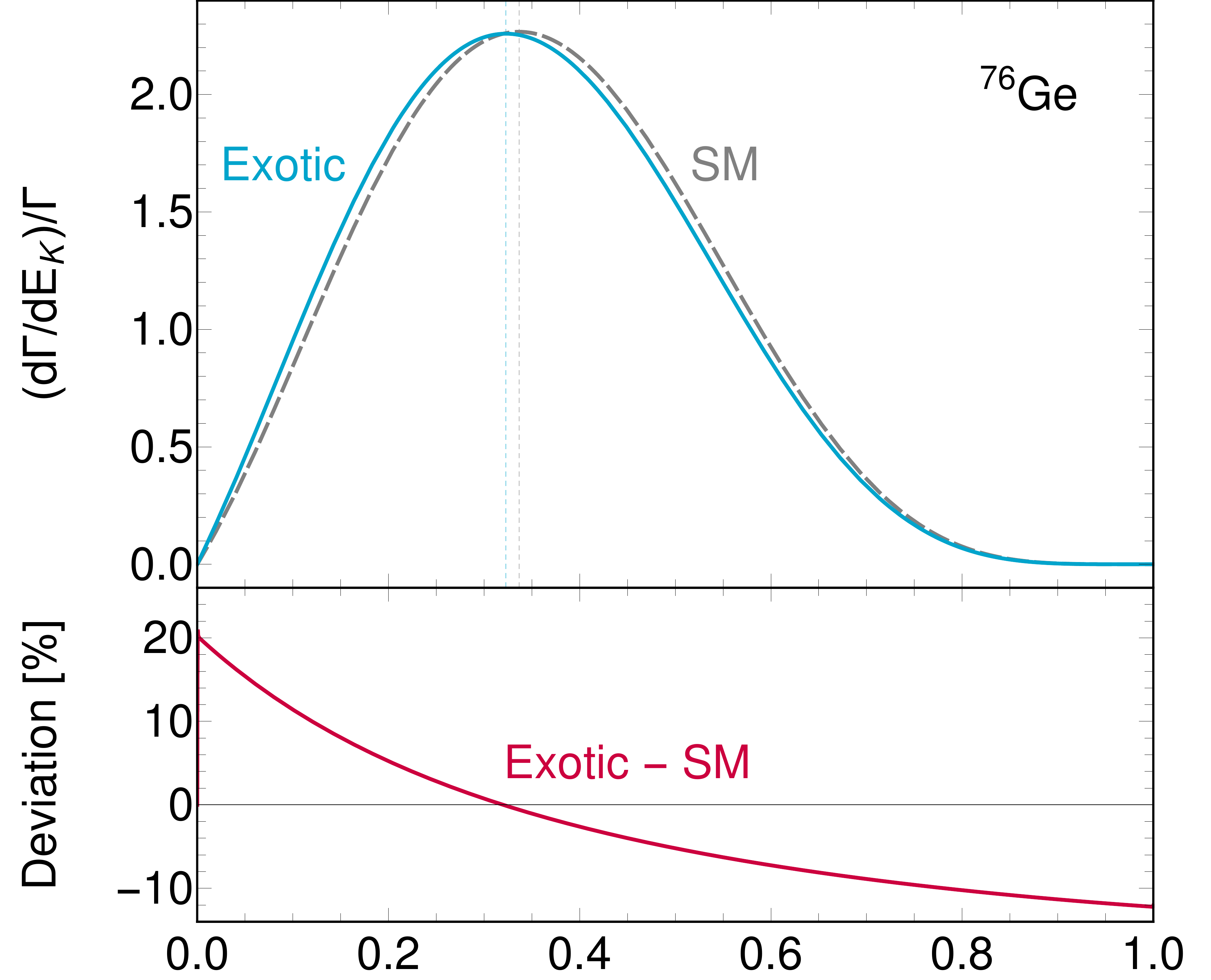}} \vspace{0.3cm}
 	\hbox{\includegraphics[width=1\textwidth]{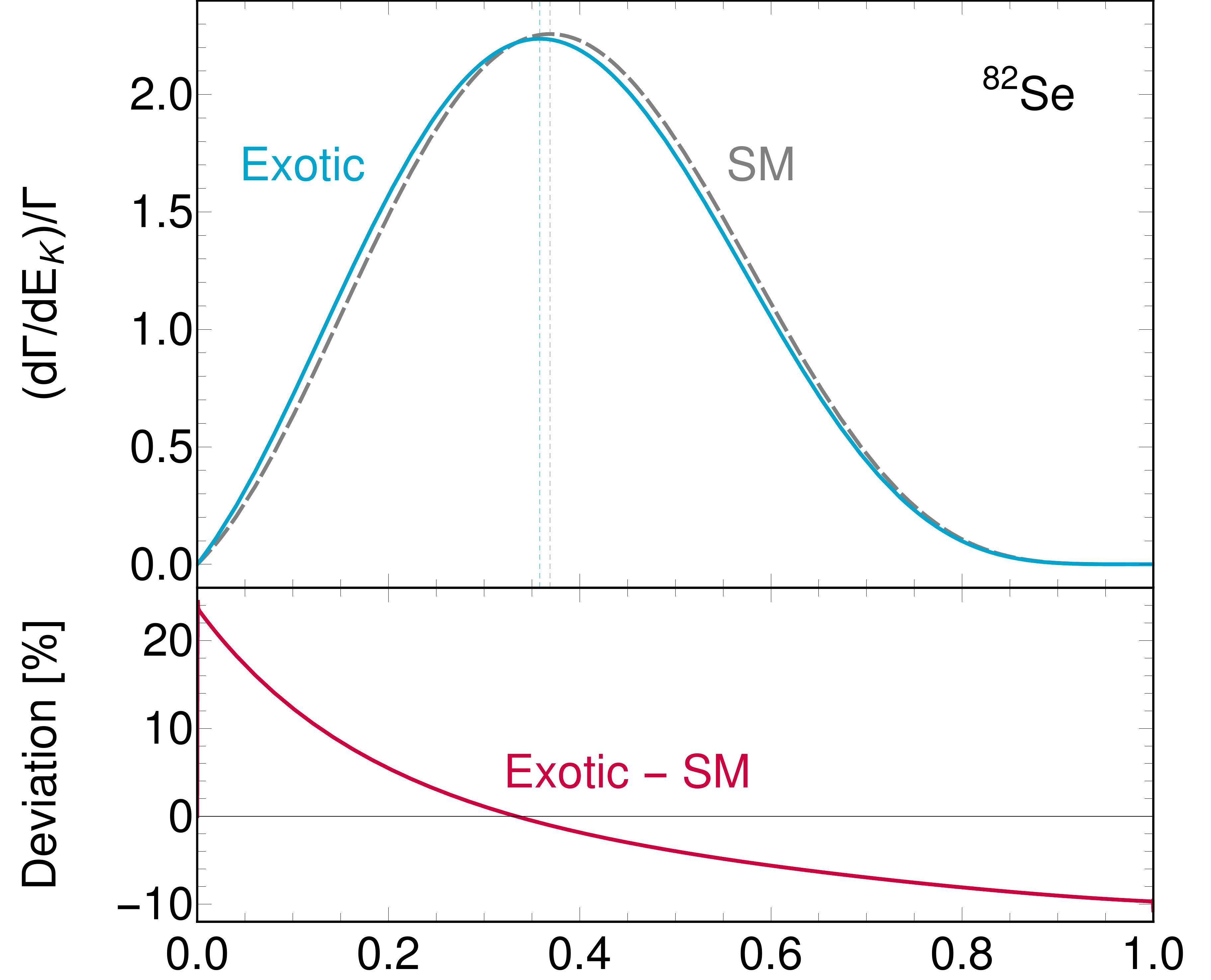}} \vspace{0.3cm}
	\hbox{\includegraphics[width=1\textwidth]{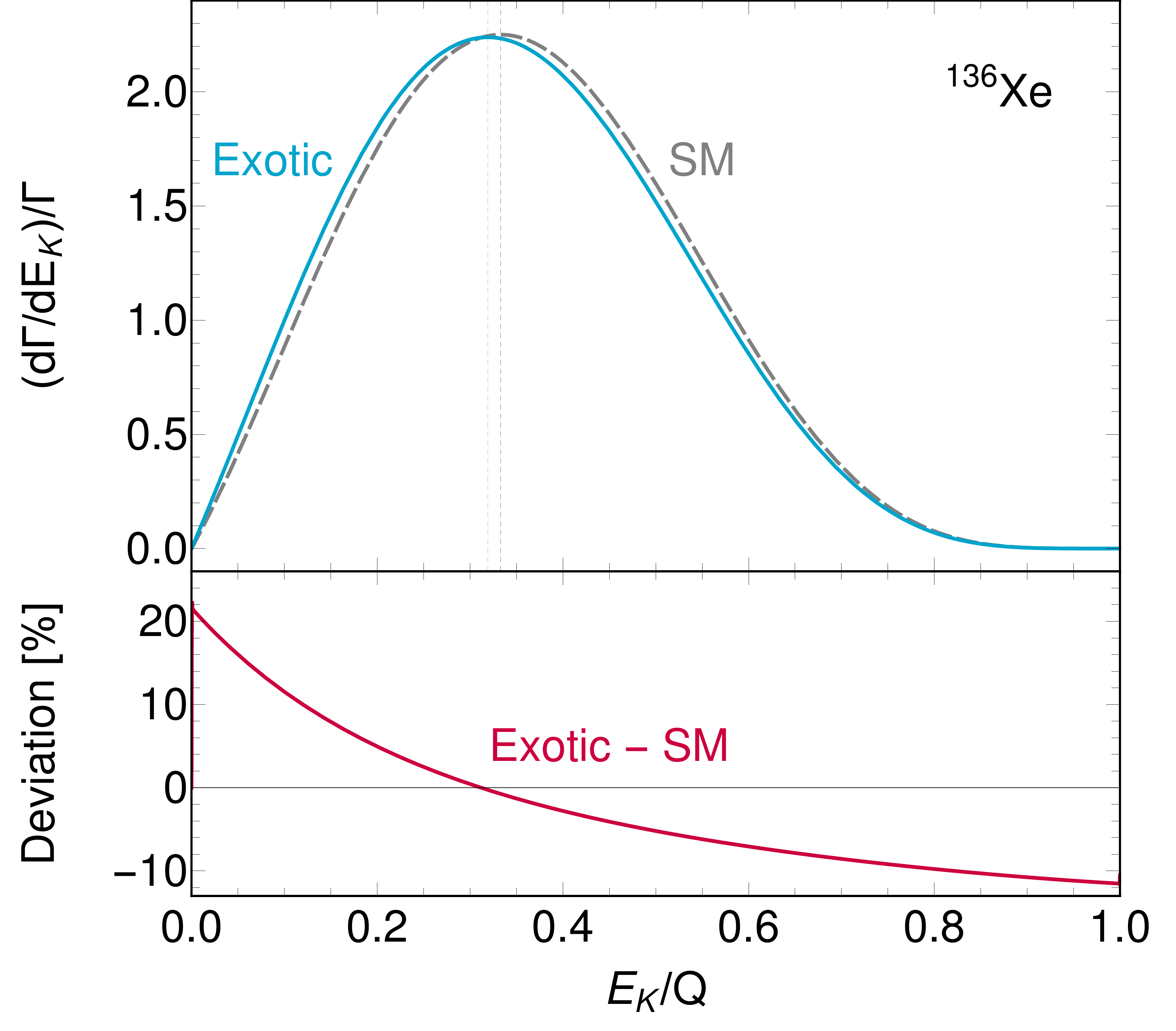}}
    }%
	\end{minipage}
	\hfill	
	%\hspace{0.333cm}
	\begin{minipage}{0.47\textwidth}
	\leavevmode
    \vbox{% baseline is at the bottom
    \offinterlineskip % disable line skip
    \hbox{\includegraphics[width=1\textwidth]{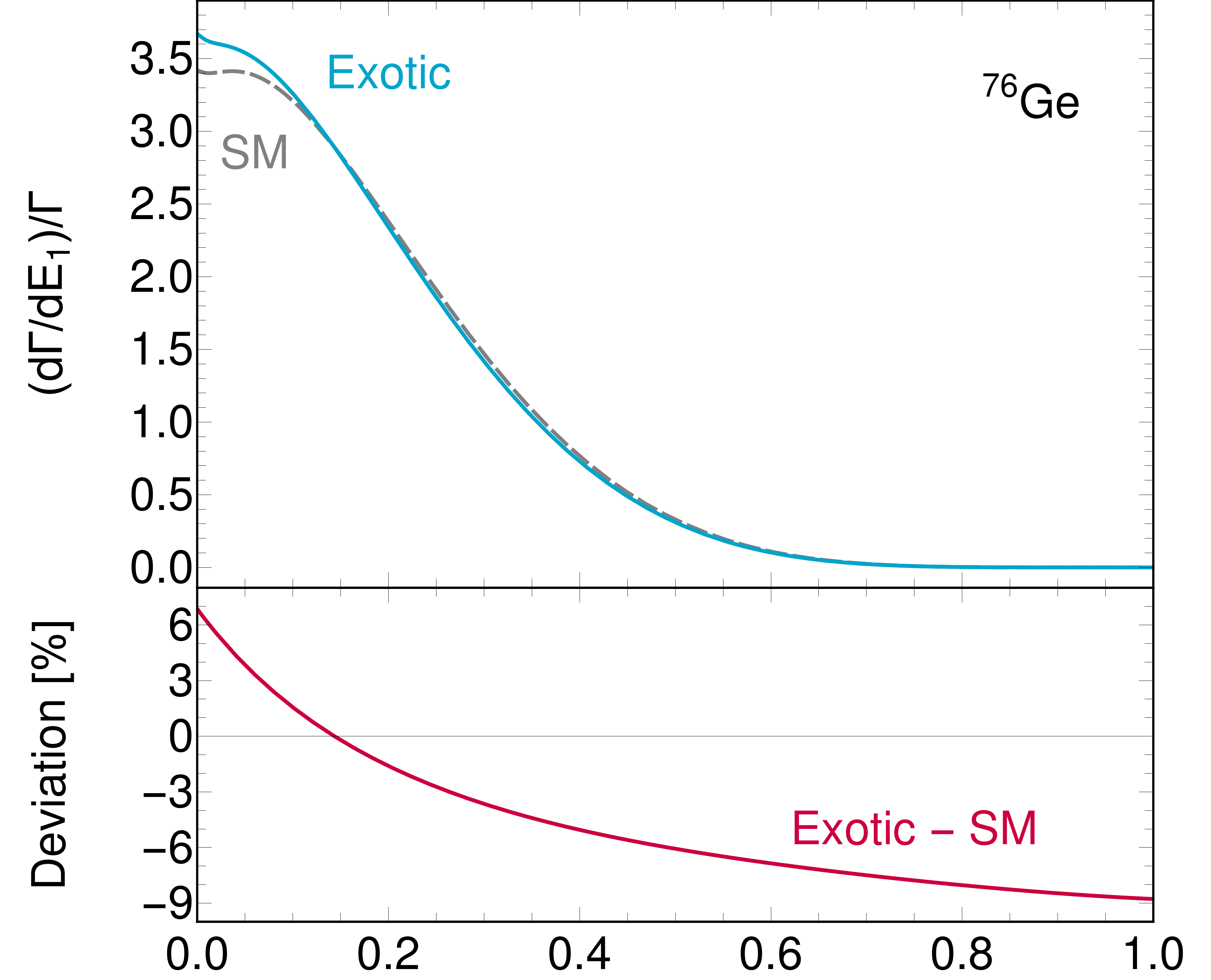}} \vspace{0.3cm}
	\hbox{\includegraphics[width=1\textwidth]{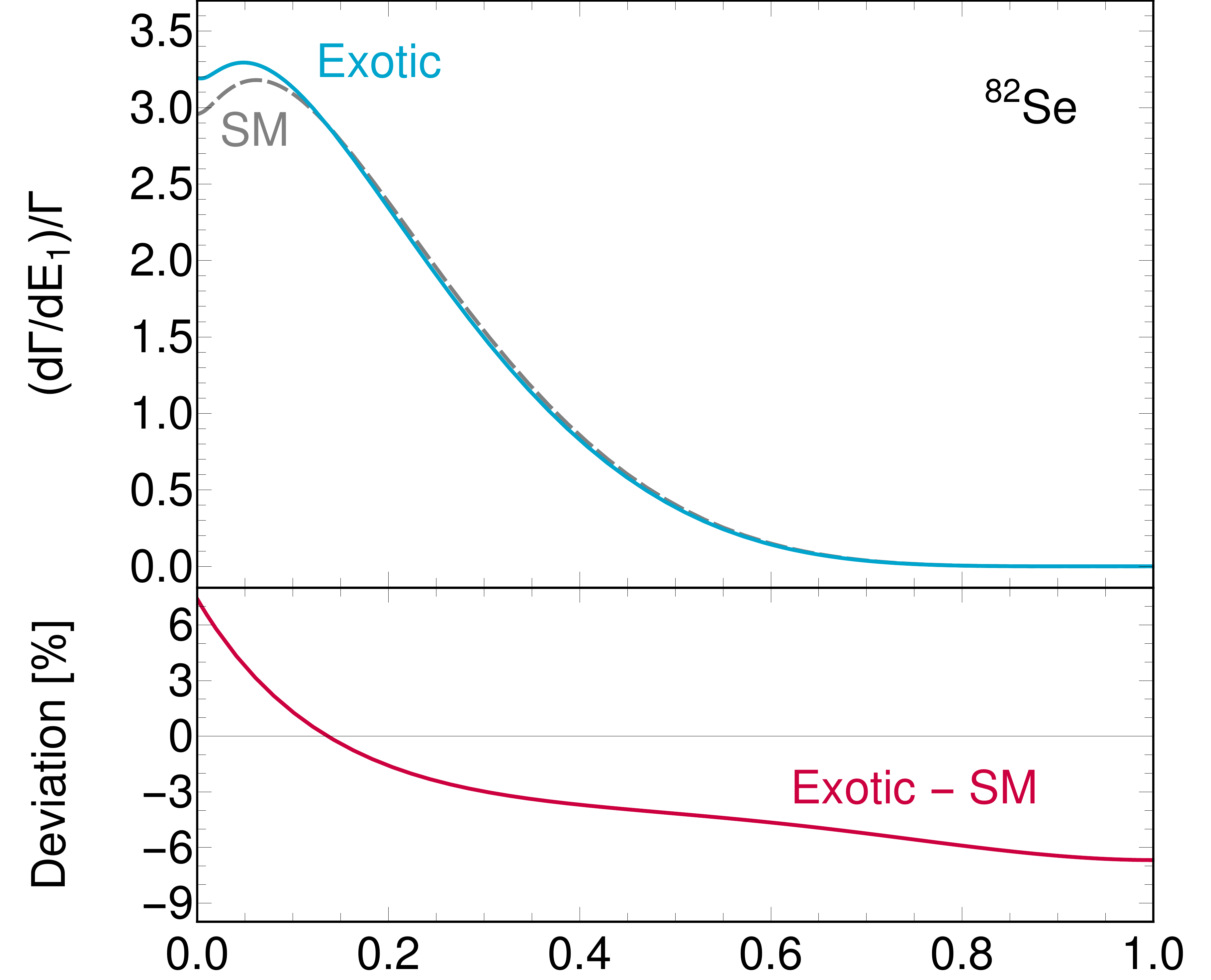}} \vspace{0.3cm}
	\hbox{\includegraphics[width=1\textwidth]{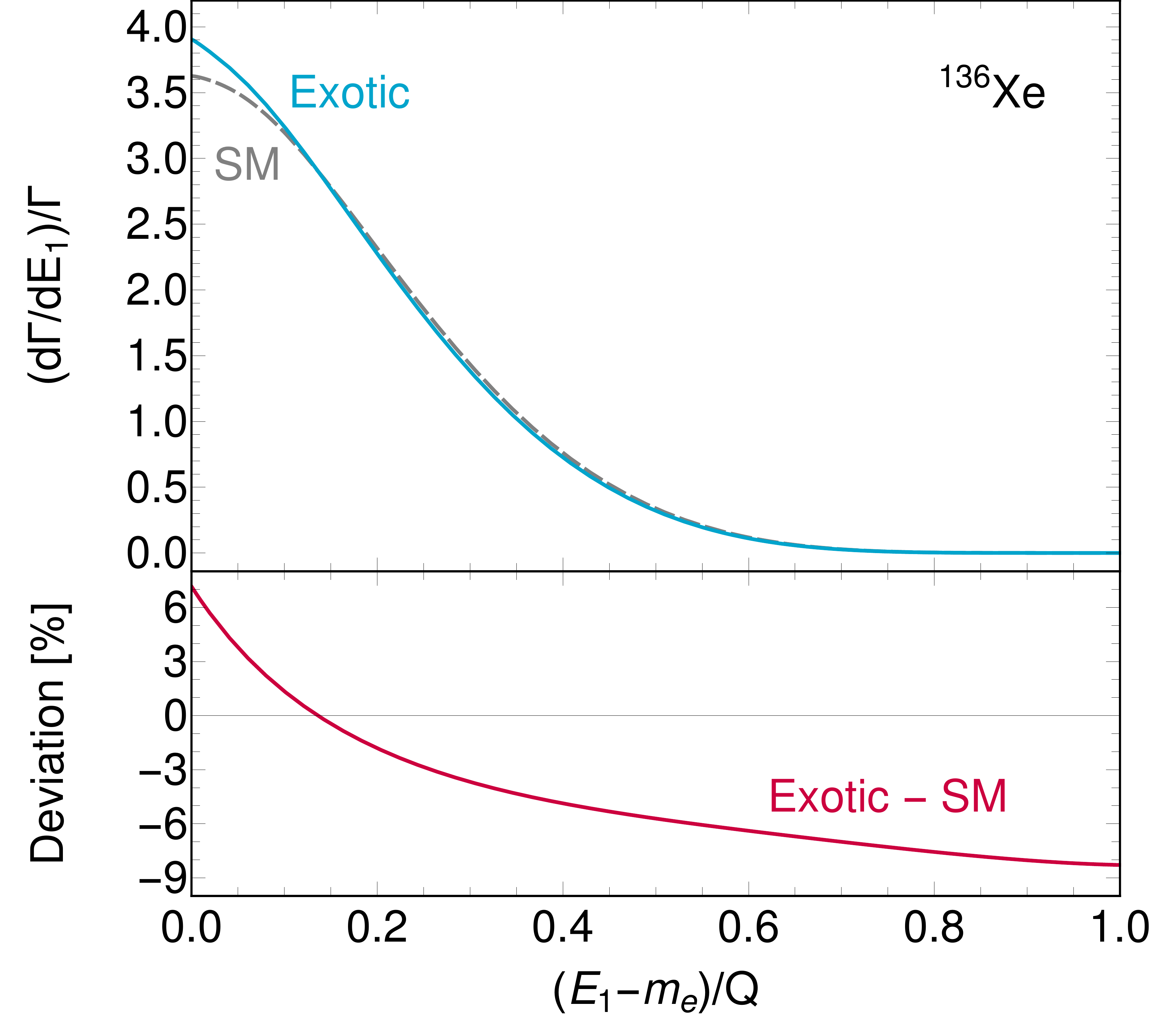}}
    }%
	\end{minipage}	
	\caption{Left: Normalized $\vvbb$ total electron kinetic energy decay distributions for SM $\vvbb$ (dashed) and the right-handed lepton current case (solid), for the isotopes ${}^{76}$Ge, ${}^{82}$Se and ${}^{136}$Xe (top to bottom). Right: Likewise, the normalized single electron energy $\vvbb$ decay distributions. The bottom panels show the relative deviation of the exotic distribution from the SM case.}
	\label{fig:otherIsosEnDistros}
\end{figure}

\paragraph*{Electron energy total and single electron energy:} 
The main observable in double beta decay experiments is the distribution with respect to the total kinetic energy of the two electrons, $d\Gamma^{2\nu}/dE_K$, $E_K = E_{e_1} + E_{e_2} - 2m_e$. In experiments where the individual electrons can be tracked and their energies measured individually, the single electron energy distribution $d\Gamma^{2\nu}/dE_{e_1}$ (by symmetry, the distribution with respect to the second electron is identical) and the double differential distribution $d\Gamma^{2\nu}/(dE_{e_1}dE_{e_2})$ are relevant as well. These distributions are calculated from Eq.~\eqref{eq:diffrate} as
\begin{align}
	\frac{d\Gamma^{2\nu}}{dE_{e_1}dE_{e_2}} &= \int_{-1}^1 d\cos\theta \frac{d\Gamma^{2\nu}}{dE_1 dE_2 d\cos\theta}, \nn\\
	\frac{d\Gamma^{2\nu}}{dE_{e_1}} &= \int_{m_e}^{Q + 2m_e - E_{e_1}}dE_{e_2}
	\frac{d\Gamma^{2\nu}}{dE_{e_1}dE_{e_2}} \nn\\
	\frac{d\Gamma^{2\nu}}{dE_K} &= 
	\int_{m_e}^{Q+m_e}dE_{e_1}dE_{e_2} \delta(E_K - E_{e_1} - E_{e_2} + 2m_e) \frac{d\Gamma^{2\nu}}{dE_{e_1}dE_{e_2}}, 
\end{align} 
The former is plotted in Fig.~\ref{fig:enDistros} and the latter two in Fig.~\ref{fig:EnergyDistros1D} of the main text, for both the SM and the exotic contribution in ${}^{100}$Mo. The total kinetic energy and single electron energy distributions for other isotopes, namely for ${}^{76}$Ge, ${}^{82}$Se and ${}^{136}$Xe, are depicted in Fig.~\ref{fig:otherIsosEnDistros}. In all the figures, the distributions are plotted with respect to the kinetic energies $E_{e_i} - m_e$ rather than the total energies $E_{e_i}$.

\begin{figure}[t]
	\centering
	\includegraphics[width=0.49\textwidth]{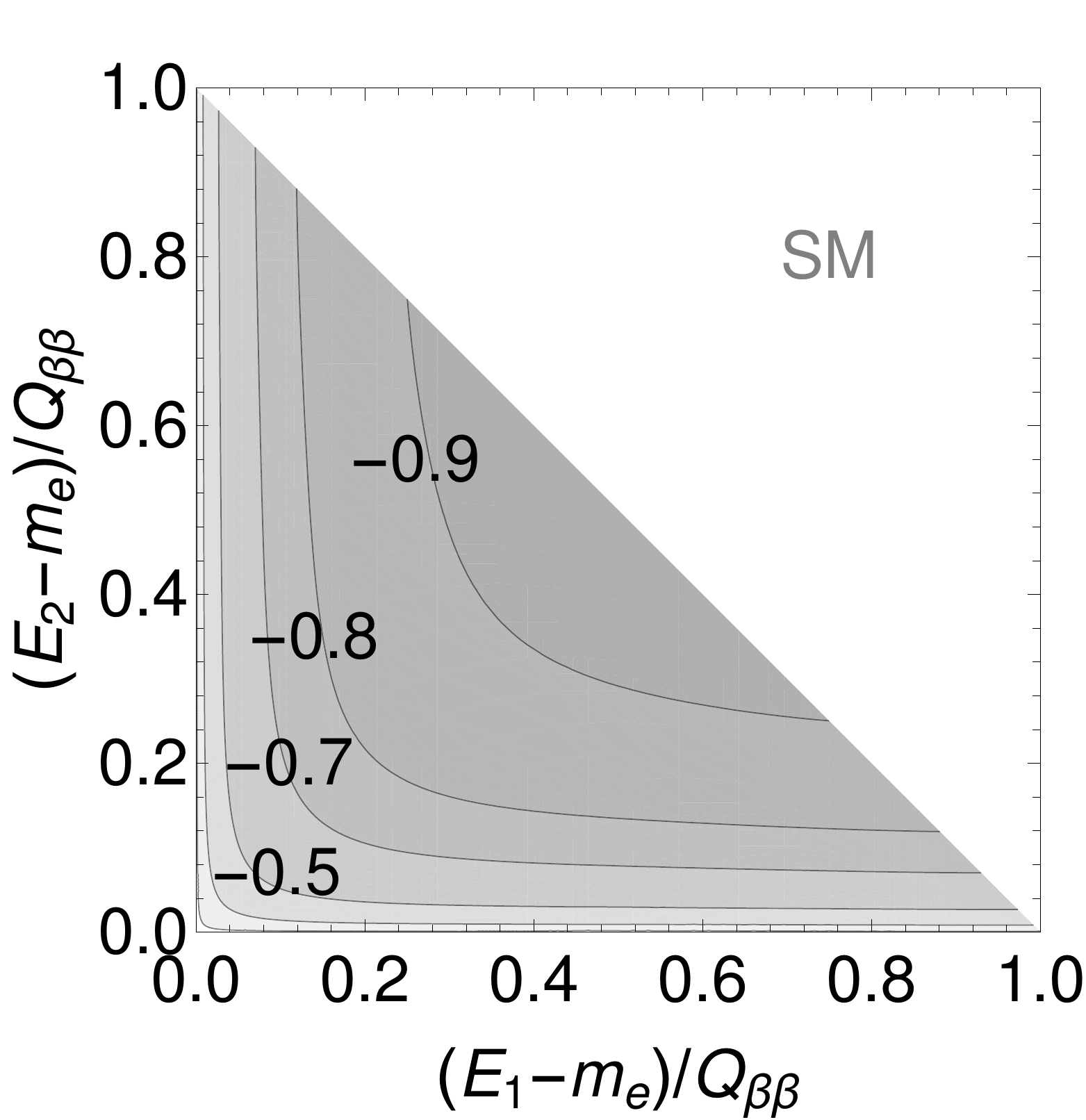}
	\hfill
	\includegraphics[width=0.49\textwidth]{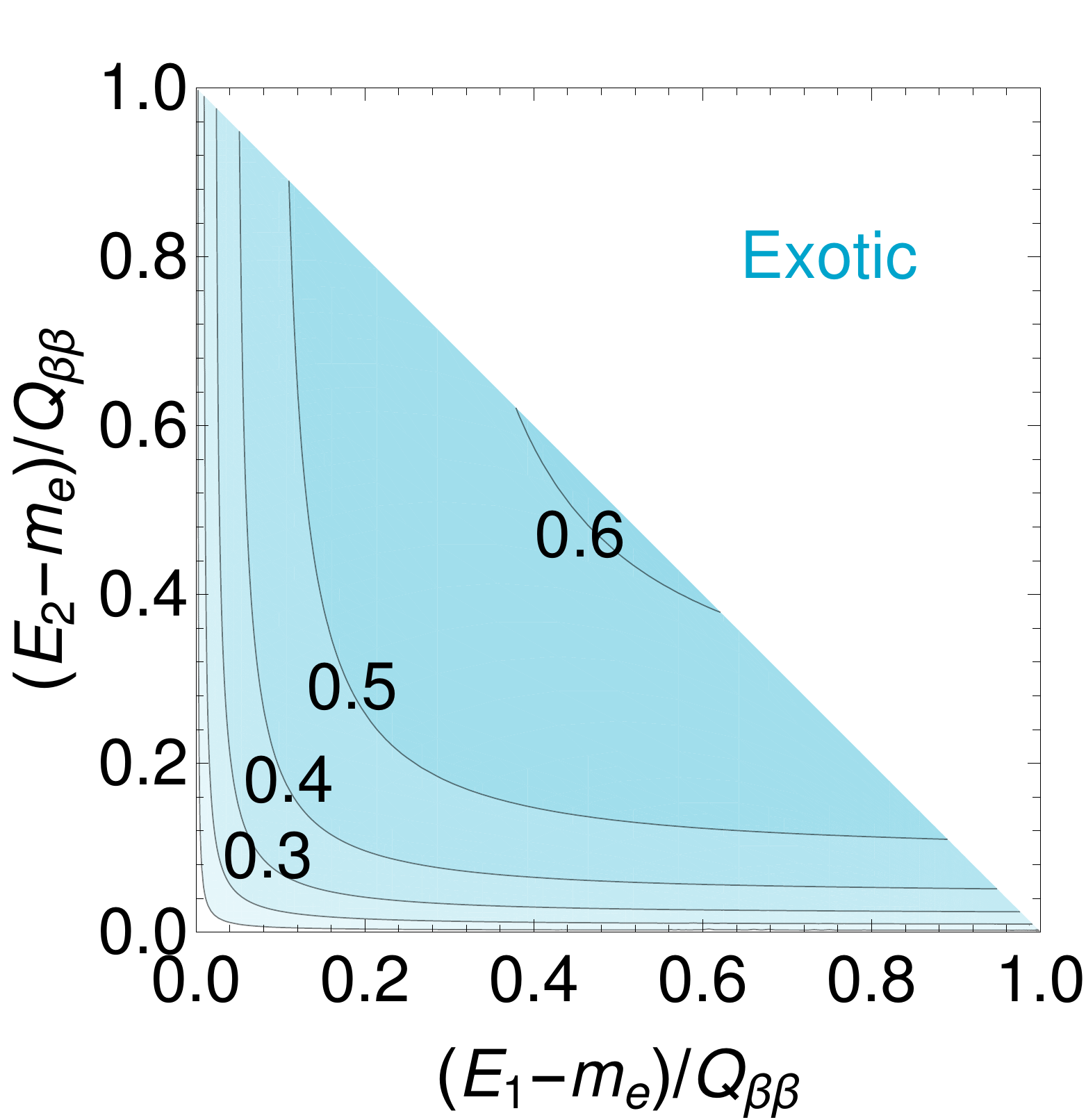}
	\caption{Angular correlation $\kappa^{2\nu}$ as a function of the electron kinetic energies for SM $\vvbb$ decay (left) and for the exotic scenario incorporating a right-handed lepton current (right). Both plots are for $^{100}$Mo and the energies are normalized to the $Q$ value.}
	\label{fig:angCorrs}
\end{figure}

\paragraph*{Energy dependent angular correlation:} One of the key consequences of right-handed lepton currents is the modification of the angular correlation between the electrons. In full generality, this is encoded in the energy-dependent angular correlation $\kappa^{2\nu}(E_{e_1}, E_{e_2})$ defined by
\begin{align}
	\kappa^{2\nu}(E_{e_1}, E_{e_2}) = 
	\frac{{B}^{2\nu}}{{A}^{2\nu}} \frac{p_{e_1} p_{e_2}}{E_{e_1} E_{e_2}} 
	= \frac{{B}^{2\nu}_{\text{SM}} 
	+ 2\epsilon_{LR} {B}^{2\nu}_{\epsilon\text{SM}} 
	+ \epsilon_{LR}^2 {B}^{2\nu}_{\epsilon}}{{A}^{2\nu}_{\text{SM}} + 2\epsilon_{LR} {A}^{2\nu}_{\epsilon\text{SM}} + \epsilon_{LR}^2 {A}^{2\nu}_{\epsilon}} \frac{p_{e_1} p_{e_2}}{E_{e_1} E_{e_2}},
\end{align}
with $A^{2\nu}_{\text{SM}}$, $A^{2\nu}_{\epsilon\text{SM}}$, $A^{2\nu}_{\epsilon}$, $B^{2\nu}_{\text{SM}}$, $B^{2\nu}_{\epsilon\text{SM}}$, $B^{2\nu}_{\epsilon}$ given by Eq.~\eqref{eq:ABnuint} applied on the SM, exotic-SM interference and exotic contributions. The resulting angular correlation is plotted in Fig.~\ref{fig:angCorrs} for both the SM contribution and the exotic contribution. The correlation $\kappa^{2\nu}$ is negative for all energies in the SM case, thus indicating that the electrons are preferably emitted back-to-back. On the contrary, the correlation is positive for the exotic scenario meaning that the electrons prefer to escape from the nucleus in the same direction.

\begin{figure}[t]
	\centering
	\includegraphics[width=0.55\textwidth]{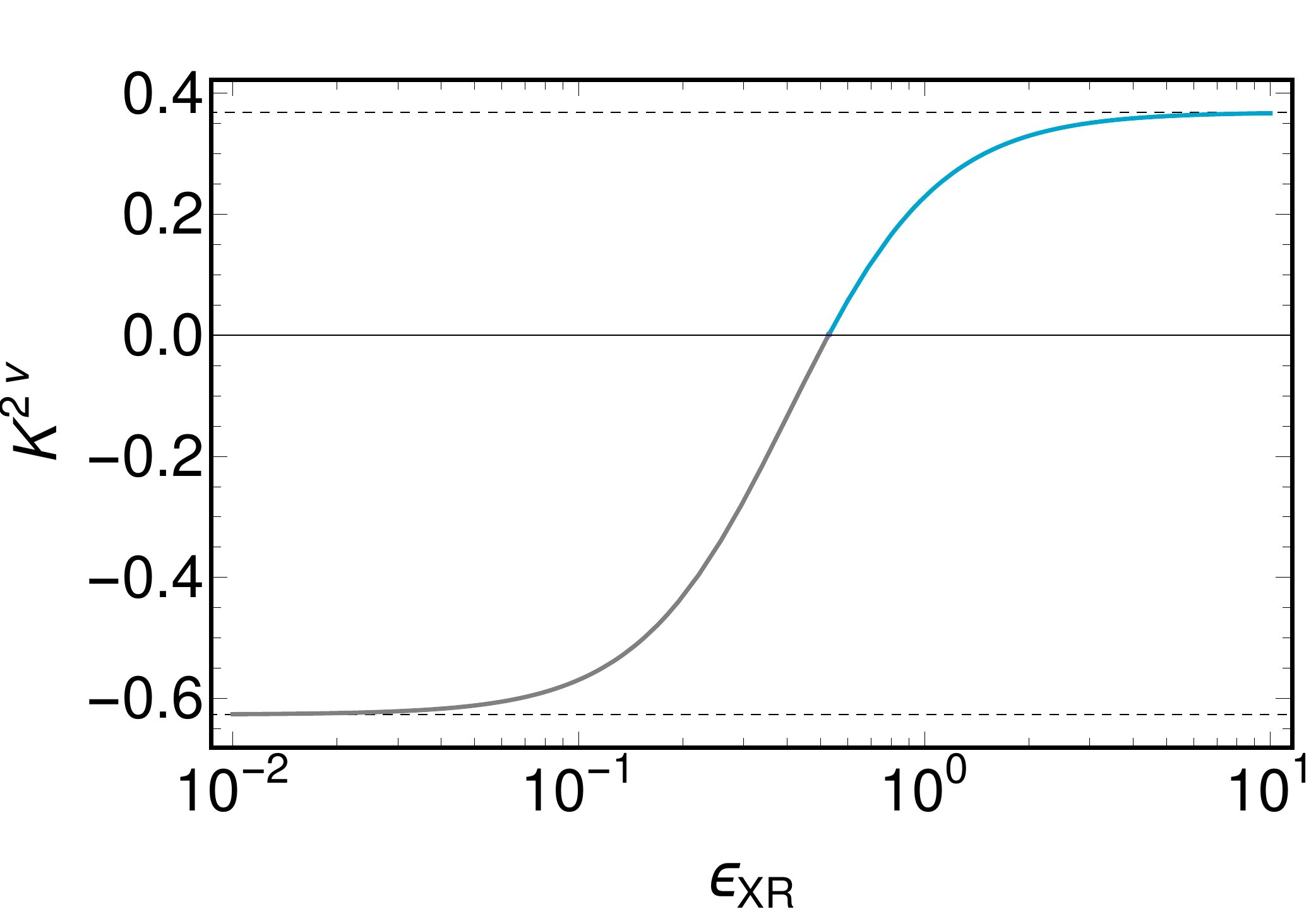}
	\caption{Angular correlation factor $K^{2\nu}$ for $^{100}$Mo as a function of the the new-physics coupling $\epsilon_{XR}$.}
	\label{fig:angFacOnEps}
\end{figure}

\paragraph*{Angular correlation factor and total decay rate:}
One can further proceed and integrate over the electron energies which yields the general form
\begin{align}
	\frac{d\Gamma^{2\nu}}{d\cos\theta} = 
	\frac{\Gamma^{2\nu}}{2}\left(1 - K^{2\nu} \cos\theta\right),
\end{align}
where $\Gamma^{2\nu}$ is the total $\vvbb$ decay rate and $K^{2\nu} = \Lambda^{2\nu}/\Gamma^{2\nu}$ is the angular correlation factor, both given by
\begin{align}
 \label{eq:gammaLambda}
	\begin{pmatrix}
    	\Gamma^{2\nu} \\
    	\Lambda^{2\nu}
	\end{pmatrix}  
 	&= \frac{c_{2\nu}}{m_e^{11}}
 	        \int_{m_e}^{E_i-E_f-m_e} dE_{e_1} p_{e_1} E_{e_1}
 	        \int_{m_e}^{E_i-E_f-E_{e_1}} dE_{e_2} p_{e_2} E_{e_2}
    \begin{pmatrix}
    	A^{2\nu} \\
    	B^{2\nu}
  	\end{pmatrix}.
\end{align}

In the case of $^{100}$Mo, the total $\vvbb$ decay rate may be approximately expressed as
\begin{align}
	\Gamma^{2\nu}_{\rm Mo} \approx \Gamma^{2\nu}_{\rm SM}(1 + 6.11\,\epsilon^2_{XR}),
\label{eq:approxrateMo}
\end{align}
where $\Gamma^{2\nu}_{\rm SM}$ is the total SM $2\nu\beta\beta$ decay rate of $^{100}$Mo. The approximated total rates for $^{76}\text{Ge}$, $^{82}\text{Se}$, $^{136}\text{Xe}$ are then given by analogous expressions,
\begin{align}
	\Gamma^{2\nu}_{\rm Ge} 
	&\approx \Gamma^{2\nu}_{\rm SM}(1 + 6.38\,\epsilon^2_{XR}), \\
	\Gamma^{2\nu}_{\rm Se} 
	&\approx \Gamma^{2\nu}_{\rm SM}(1 + 6.07\,\epsilon^2_{XR}), \\
	\Gamma^{2\nu}_{\rm Xe} 
	&\approx \Gamma^{2\nu}_{\rm SM}(1 + 6.26\,\epsilon^2_{XR}).
\end{align}
Here, $\Gamma^{2\nu}_{\rm SM}$ are again the SM $2\nu\beta\beta$ decay rates of the respective isotope.

The angular correlation factor for the SM contribution in $^{100}$Mo is $K^{2\nu}_{\text{SM}} = -0.63$ and for the exotic contribution it is $K^{2\nu}_\epsilon = +0.37$. In general, $K^{2\nu}$ as a function of $\epsilon_{XR}$ for $^{100}$Mo is plotted in Fig.~\ref{fig:angFacOnEps}. This clearly shows that admixtures of the SM and exotic contributions interpolate between the SM case ($\epsilon_{XR}=0$) and a dominant exotic case ($\epsilon_{XR}\gg 1$). For the physically relevant case where $\epsilon_{XR} \ll 1$, the factor is well approximated by
\begin{align}
	K^{2\nu}_{\text{Mo}} \approx -(0.6260\pm 0.0030) + (6.078\pm 0.017)\,\epsilon_{XR}^2.
\end{align}
Here, the uncertainties are from varying the effective axial coupling in the range $0.8 \leq g_A \leq 1.269$. The analogous equations for $^{76}\text{Ge}$, $^{82}\text{Se}$, $^{136}\text{Xe}$ read
\begin{align}
	K^{2\nu}_{\text{Ge}} &\approx -0.53 + 5.3\,\epsilon_{XR}^2, \\
	K^{2\nu}_{\text{Se}} &\approx -0.64 + 6.2\,\epsilon_{XR}^2, \\
	K^{2\nu}_{\text{Xe}} &\approx -0.57 + 5.6\,\epsilon_{XR}^2,
\end{align}
respectively, for $g_A = 1.0$. As apparent, the dependence of the correlation factor on a small exotic coupling $\epsilon_{XR}$ is similar for different isotopes.

\bibliography{references}
\end{document}